\documentclass[useAMS,usenatbib]{mn2e}
\usepackage{graphicx,rotate,url,mathptmx,lscape,times, color,subfigure}
\def\gtsim{\mathrel{\hbox{\rlap{\hbox{\lower4pt\hbox{$\sim$}}}\hbox{$>$}}}}
\def\lesssim{\mathrel{\hbox{\rlap{\hbox{\lower4pt\hbox{$\sim$}}}\hbox{$<$}}}}
\def\Msunpyr{M$_{\odot}\,$yr$^{-1}$}
\def\Msun{M$_{\odot}$}
\def\erg{{\rm\thinspace erg}}

\def\Hz{{\rm\thinspace Hz}}

\def\km{{\rm\thinspace km}}

\def\Mpc{{\rm\thinspace Mpc}}
\def\Msun{\hbox{$\rm\thinspace M_{\odot}$}}

\def\s{{\rm\thinspace s}}
\def\ps{{\rm\thinspace s^{-1}}}

\def\yr{{\rm\thinspace yr}}

\def\simless{\mathbin{\lower 3pt\hbox
	{$\,\rlap{\raise 5pt\hbox{$\char'074$}}\mathchar"7218\,$}}} 
\def\simgreat{\mathbin{\lower 3pt\hbox
	{$\,\rlap{\raise 5pt\hbox{$\char'076$}}\mathchar"7218\,$}}} 

\def\ergps{\hbox{$\erg\s^{-1}\,$}}
\def\ergpsphz{\hbox{$\ergps\Hz^{-1}\,$}}

\def\kmps{\hbox{$\km\ps\,$}}
\def\kmpspMpc{\hbox{$\kmps\Mpc^{-1}\,$}}

\def\Msunpyr{\hbox{$\Msun\yr^{-1}\,$}}

\def\h0{\hbox{{\rm H}$^0$}}
\DeclareMathAlphabet{\vib}{OML}{cmm}{m}{it}

\def\h{$H_{\rm 160}$}

\def\prob{$\mathcal P$}

\def\clustername{{Cl\,0218.3$-$0510}}
\begin{document}
\title[The structure of a protocluster]{The structure and evolution of a forming galaxy cluster at $z=1.62$}
\author[N.\,A.\,Hatch et al.]
       {\parbox[]{6.0in}
       {N.\,A.\,Hatch$^{1}$\thanks{E-mail: nina.hatch@nottingham.ac.uk},~S.\,I.\,Muldrew$^2$, E.\,A.\,Cooke$^1$, W.\,G.\,Hartley$^3$, O.\,Almaini$^1$, C.\,J.\,Simpson, C.\,J.\,Conselice$^1$.
       \\        \footnotesize
        $^1$School of Physics and Astronomy, University of Nottingham, University Park, Nottingham NG7 2RD, UK\\  
	$^2$Department of Physics and Astronomy, University of Leicester, University Road, Leicester, LE1 7RH, UK\\
        $^3$ETH Z\"urich, Institut f\"ur Astronomie, HIT J 11.3, Wolfgang-Pauli-Str. 27, 8093, Z\"urich, Switzerland.
    }}
 \date{}
\pubyear{}
\maketitle

\label{firstpage}
\begin{abstract}
We present a comprehensive picture of the \clustername\ protocluster at $z=1.623$ across 10 co-moving Mpc. Using filters that tightly bracket the Balmer and 4000\AA\ breaks of the protocluster galaxies we obtain precise photometric redshifts resulting in a protocluster galaxy sample that is $89\pm5$\% complete and has a contamination of only $12\pm5$\%. Both star forming and quiescent protocluster galaxies are located allowing us to map the structure of the forming cluster for the first time. The protocluster contains $6$ galaxy groups, the largest of which is the nascent cluster. Only a small minority of the protocluster galaxies are in the nascent cluster (11\%) or in the other galaxy groups (22\%), as most protocluster galaxies reside between the groups. Unobscured star forming galaxies predominantly reside between the protocluster's groups, whereas red galaxies make up a large fraction of the groups' galactic content, so observing the protocluster through only one of these types of galaxies results in a biased view of the protocluster's structure. The structure of the protocluster reveals how much mass is available for the future growth of the cluster and we use the Millennium Simulation, scaled to a Planck cosmology, to predict that \clustername\ will evolve into a $2.7^{+3.9}_{-1.7}\times 10^{14}$\Msun\ cluster by the present day.  
\end{abstract}


\section{Introduction}

Galaxy clusters are unique laboratories to study galaxy formation. Distant clusters are the statistical ancestors of present-day clusters, so we can study the processes that drive galaxy evolution by comparing the galaxies within clusters at low and high redshifts.  Distant clusters contain more blue, spiral galaxies with higher star formation rates than cluster members today  \citep{Poggianti2009a}, as well as a large fraction of post-starburst galaxies \citep{Poggianti2009b,Muzzin2012} and a lack of low-mass red galaxies \citep[e.g.,][]{DeLucia2004b,Rudnick2012}, all of which implies strong galaxy evolution.

The progenitors of present-day clusters are called \lq protoclusters\rq. These are agglomerations of galaxies and groups that will merge to form a cluster by the present day.  These bound structures are very extended, stretching up to 50\,co-moving\,Mpc in diameter \citep{Muldrew2015}. 
Following the definition of a cluster as a virialized structure, we therefore refer to the most massive halo of the protocluster as the high-redshift cluster. These nascent clusters are relatively compact objects, typically $R_{500}\sim0.5$\,Mpc or less  \citep[e.g.][]{Fassbender2014}, and contain only a small fraction of the galaxies that will eventually form the present-day cluster. Cosmological simulations reveal that the majority of galaxies in clusters at $z=0$ did not reside in the main halo at $z>1$, but rather in the extended protocluster \citep{Muldrew2015}. It is therefore essential to study the protocluster, and not simply the main halo, to trace the evolution of cluster galaxies to the highest redshifts.

By tracing the evolution of all protocluster galaxies, we take into account the variety of environments that a galaxy experiences as the cluster collapses. Galaxies that end up in clusters start in low density filaments and migrate to dense groups due to gravity. Each environment imprints itself on the properties of the galaxies, so the final result is the sum of all the environments in which the galaxy has ever lived. To trace the evolution of galaxies that end up in the cluster core, we must identify all of their ancestors -- those that reside in dense environments at high redshift, and those that migrate there at later times. 

Tens of protoclusters and $z>1.5$ clusters have been spectroscopically confirmed to date 
and hundreds of protocluster candidates are known \citep[e.g.][]{Wylezalek2013,Planck2015}. The handful of $z>1.5$ protoclusters that have been studied in detail reveal a strongly star-forming galaxy population, and accelerated mass growth compared to field galaxies \citep{Steidel2005,Hatch2011b,Cooke2014}, which supports a picture of accelerated galaxy formation before the cluster has assembled. 
But in comparison to lower-redshift clusters, our understanding of $z>1.5$ clusters and protoclusters is woefully poor. The intrinsic variety of galaxy protoclusters means that full census has not been reached on many issues. For example, it is not clear whether star formation is enhanced in protoclusters \citep{Tran2010} or suppressed \citep{Quadri2012}; whether galaxies are larger in size within protoclusters or not (\citealt{Papovich2012} versus \citealt{Newman2014}); or whether the member galaxies are metal deficient \citep{Valentino2015}, metal enriched \citep{Shimakawa2015}, or no different to the field \citep{Kacprzak2015,Tran2015}.

The primary challenge to studying protoclusters is obtaining a sample of protocluster galaxies that is clean from field galaxy contaminants and yet complete enough that all types of protocluster galaxies are detected to sufficiently low masses. Clean samples of protocluster galaxies are essential to isolate weak environmental trends and robustly compare properties of galaxies at different redshifts. A complete sample is required to trace galaxies as they evolve from star forming to quiescent objects.

The drive to ensure clean samples have led to the rise of narrow-band imaging surveys that isolate protocluster H$\alpha$ emitters (e.g.\,MAHALO; \citealt{Hayashi2012}) and near-infrared spectroscopic surveys (e.g.\,ZFIRE; \citealt{Yuan2014}). These surveys are remarkably clean, but they miss galaxies without AGN and those with low star formation rates. By contrast, photometric redshifts are able to identify all types of protocluster galaxies, but redshifts from broadband photometry are imprecise, with a typical accuracy no better than $\Delta z/(1+z)\sim0.03$ at $z>1$, so the contamination by field galaxies is high. For example, the broadband photometric redshift catalogue of the XMM--LSS\,J02182-05102 protocluster at $z=1.6233$ by \citet{Papovich2012} has a $32$\% contamination rate when compared with the spectroscopic survey of \citet{Tran2015}.

In addition to the problem of identifying protocluster members, is the challenge of interpreting observations of protoclusters. To use protoclusters to study galaxy evolution we must place them into evolutionary sequences of statistical ancestors and descendants. But we currently lack the tools to convert observations of protoclusters into meaningful parameters that characterise their present evolutionary state and predict their future growth and $z=0$ mass.  We must therefore derive tools that enable us to accurately estimate a protocluster's $z=0$ mass and rate of growth from an observation at a single snapshot in time. 

Recently progress towards this has been made by \citet{Chiang2013} who demonstrate there is a strong correlation between the descendant $z=0$ cluster mass and the mass overdensity of a protocluster over large volumes (15-25 comoving Mpc [cMpc] diameter). The uncertainty of the descendant $z=0$ cluster mass for each protocluster is very large, due to both intrinsic scatter in the relation and the inherent  problems in converting the observed projected galaxy overdensity into the true 3D mass overdensity \citep{Shattow2013}, but such methods can be successful in estimating the statistically likely mass of a large sample of protoclusters.

Here we present a study on the XMM--LSS\,J02182-05102 protocluster at $z=1.6233$ (referred to as \clustername\ from now on) that demonstrates how we can solve these issues and use protoclusters as laboratories to study galaxies evolve over cosmological times. 

The redshift of  \clustername\ is known, so we use special filters that span the Balmer and 4000\AA\ breaks to obtain high precision photometric redshifts. Using this precision we select a \lq Goldilocks\rq\ sample of protocluster galaxies: a sample that is both clean enough and complete enough that we can robustly study the properties of the protocluster galaxies and trace the wide-field structure of the protocluster. Using cosmological simulations to identify protoclusters with the same wide-field structure we demonstrate that we can accurately determine its ultimate mass, and its likely growth rate across time. As such we demonstrate that we can place the protocluster in cosmological context, and with similar data on (proto)clusters at other redshifts we can locate its statistical progenitor protocluster and descendant cluster.

In this paper we present the method for selecting a clean and complete sample of protocluster galaxies (Section 2). We estimate the completeness and contamination of this method, map the structure of the protocluster and investigate how observing only one type of galaxy can bias our view of the protocluster (Section 3). Finally we explore what the structure of the protocluster can reveal about its future growth (Section 4). In the accompanying paper we present properties of the protocluster galaxies as a function of their environment (Hatch et al.\,in prep). We use AB magnitudes throughout and a $\Lambda$CDM flat cosmology with $\Omega_M=0.315$, $\Omega_\Lambda=0.685$ and $H_0=67.3$ \kmpspMpc\ \citep{Planckcosmology2014}. In this cosmology, \clustername\ at $z=1.6233$ has a scale of 8.71\,kpc\,arcsec$^{-1}$.

\section{Method}
\label{method}
\subsection{Data}
\subsubsection{Broad-band Photometry}

\begin{figure*}
\includegraphics[width=1\columnwidth]{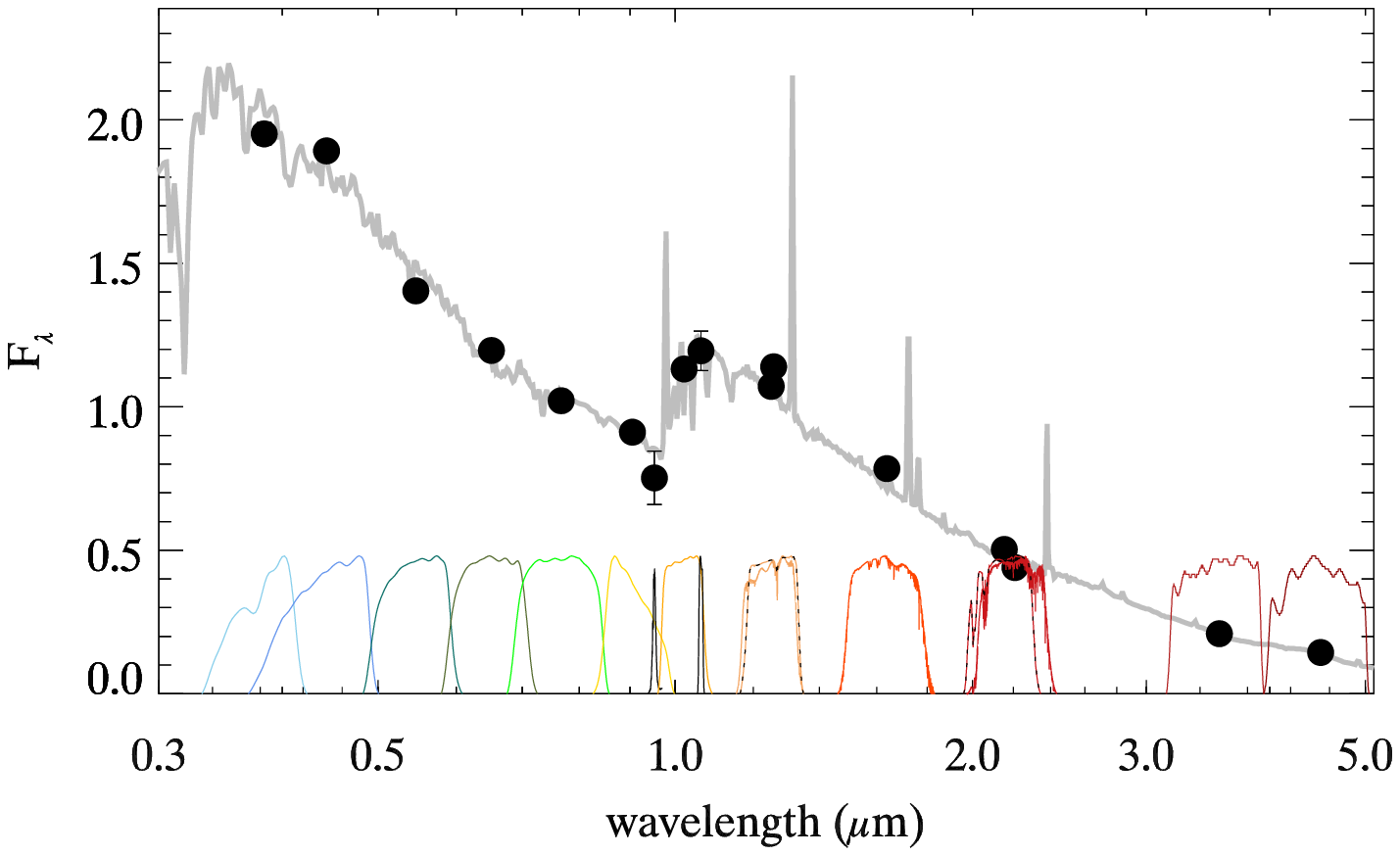}
\includegraphics[width=1\columnwidth]{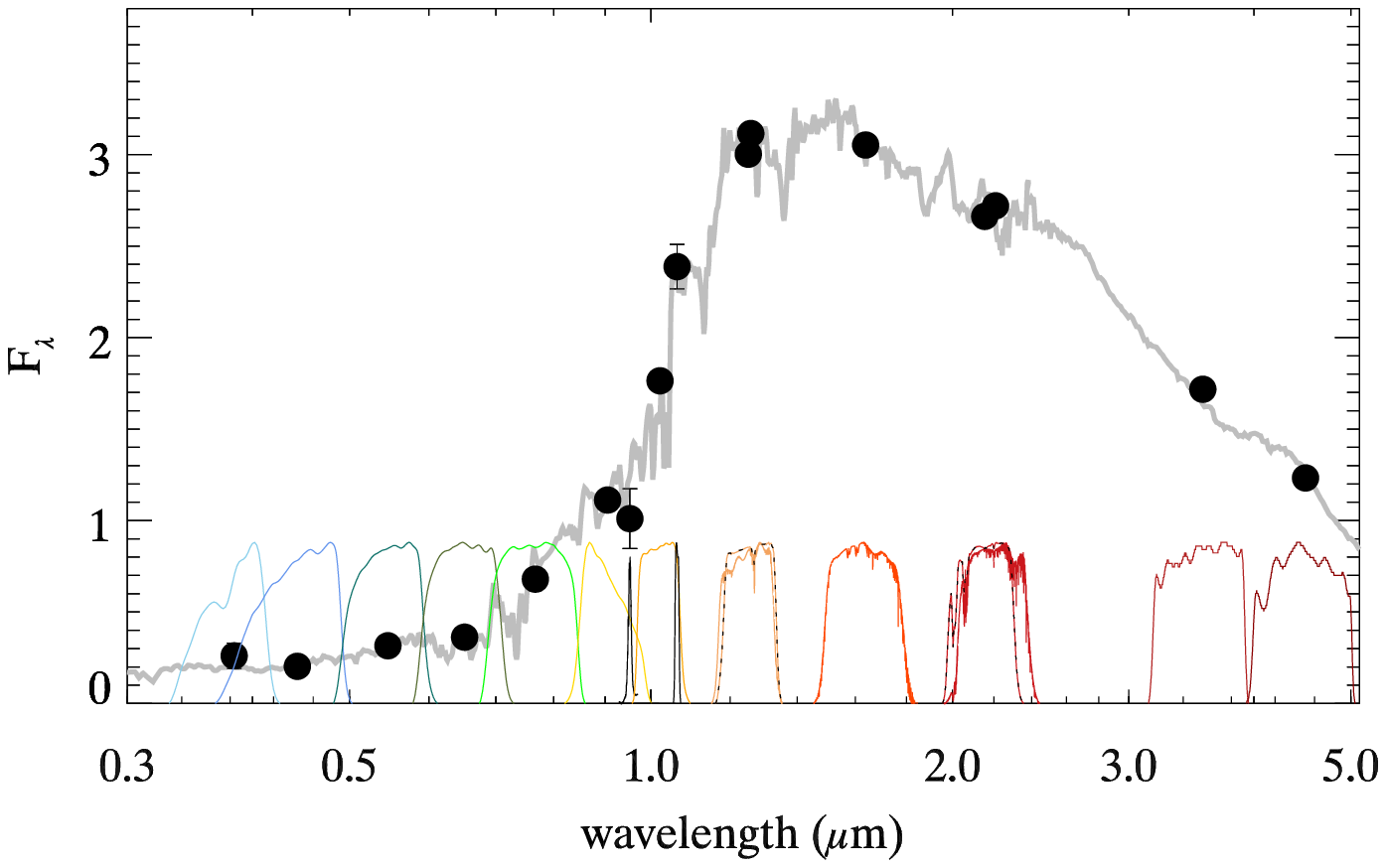}
\caption{\label{SED_examples}Examples of the SED coverage of a star forming (left) and passive (right) protocluster galaxy. The solid black circles plot the measured photometry with 1$\sigma$ uncertainties. On the bottom are the filter transmission curves of all bands used to derive photometric redshifts and galaxy properties. From left to right: $U$, $B$, $V$,$R$, $i^\prime$, $z^\prime$, [S{\sc iii}]$+65$, $Y$, NB1.06,  $J_{WFCAM}$, $J_{HAWK-I}$, H, $Ks$, $K$, IRAC1, IRAC2. The grey line shows the best-fit galaxy template assigned to the photometry by the SED fitting code {\sc fast}. Multiple wavelength coverage around $1\micron$ means the Balmer and 4000\AA\ breaks can be differentiated in the protocluster galaxies allowing precise photometric redshifts and accurate SED fitting.}
\end{figure*}

\clustername\ is covered by several deep optical and infrared surveys: the near-infrared UKIDSS Ultra Deep Survey (UDS; Almaini et al., in prep.), the optical {\it Subaru}/{\it XMM-Newton} Deep Survey (SXDS; \citealt{Furusawa2008}) and the {\it Spitzer} Ultra Deep Survey data (SpUDS;  PID 40021, P.I. J. Dunlop). We use photometry from these surveys compiled by \citet{Simpson2012} and \citet{Hartley2013} (hereafter referred to as H13), who combined $U$--band data from the Canada-France-Hawaii Telescope with $BVRi^\prime z^\prime$ optical photometry from the SXDS, {\it JHK} photometry from the eighth data release of the UDS, and SpUDS to create a  $K$--selected  $UBVRi^{\prime}z^{\prime}JHK$[3.6][4.5] catalogue (see Table \ref{table_data} for details of each image). Photometry was measured in 2\,arcsec diameter apertures because the protocluster galaxies are often very close to one another.  Aperture corrections were applied to the $U$, [3.6] and [4.5] data to account for the large difference in the point spread function (PSF) between these images and the rest of the optical and infrared data.

Additional deep broad-band $Y$, $J$, and $Ks$ images were obtained as part of ESO programme 386.A-0514 (P.I.~Tran) using HAWK-I \citep{Kissler-Patig2008} on the ESO Very Large Telescope (VLT). HAWK-I is a near-IR camera comprising four Hawaii-2 $2048\times2048$ pixel detectors separated by a gap of $\sim15$\,arcsec. The camera spans $7.5 \times 7.5$\,arcmin with a pixel scale of 0.106\,arcsec\,pixel$^{-1}$.  The data were reduced using standard near-infrared reduction techniques with the ESO {\sc mvm} software \citep{Vandame2004}. Flux calibration was achieved using the UDS and WFCAM to HAWK-I conversions derived for VIRCAM\footnote{http://casu.ast.cam.ac.uk/surveys-projects/vista/technical/photometric-properties}
, which has almost identical filters to HAWK-I. The $K-$selected catalogue of H13 was used for the basis of the astrometry, so the resulting images are matched to the H13 catalogue to within 0.1\,pixel~(0.01\,arcsec). See Table \ref{table_data} for details regarding final exposure time, seeing and image depth. 

\subsubsection{Narrow-band images bracketing the Balmer and 4000\AA\ breaks}
The strongest spectral features for most of the protocluster galaxies are the Balmer (3686\AA) and 4000\AA\ breaks. The Balmer break is strongest for galaxies which are still forming some stars, whilst the 4000\AA\ break is most prominent in passively evolving galaxies (see Fig.\,\ref{SED_examples}). High precision photometric redshifts can be achieved if these features are well-sampled. We therefore obtained narrow-band images of the protocluster at 9530\AA\ and 10600\AA\ using the ESO/VLT FORS $[$S{\sc iii}$]+$65 and HAWK-I 1.06$\micron$ (NB1.06) filters, which cover rest-frame 3630\AA\ and 4040\AA, respectively, for galaxies in the protocluster at $z\sim1.623$. The Balmer break is tightly bracketed by the FORS $[$S{\sc iii}$]+$65 and $Y$ filters, whilst the 4000\AA\ break is sampled by the $Y$, NB1.06 and $J$ filters (see Fig\,\ref{SED_examples}). The images were reduced using the publicly available {\sc theli} software \citep{Erben2005,Schirmer2013}. Flux calibration was achieved by linearly interpolating the photometry of the H13 catalogue, and this catalogue was again used as the basis for the astrometric calibration. See Table \ref{table_data} for details regarding the exposure time, seeing and image depth. 

\begin{table*}
\begin{tabular}{|l|c|c|c|c|c|l|}
\hline
Filter&Central wavelength &  Instrument/ & Reference &Exposure time& Depth (2\arcsec aperture)& PSF FWHM  \\
&(\AA)& Telescope &  & (mins) &  (5$\sigma$)& (arcsec)  \\
 \hline
U&	3835&		Megacam/CFHT		& 	H13 & 	350	&25.76	& 1.03	\\
B&	4435&		Suprime-Cam/SUBARU	& 	\citet{Furusawa2008}&	345	&27.6 	& 0.80	\\
V&	5462&		Suprime-Cam/SUBARU	& 	\citet{Furusawa2008}&	319	&27.2 	& 0.72	\\
R&	6515&		Suprime-Cam/SUBARU	& 	\citet{Furusawa2008}&	248	&27.0 	& 0.76 	\\
i$^\prime$&	7666&		Suprime-Cam/SUBARU	& 	\citet{Furusawa2008}&	647	&27.0 & 0.78	\\
z$^\prime$&	9052&		Suprime-Cam/SUBARU	& 	\citet{Furusawa2008}&	217	&26.0 	& 0.70	\\
$[$S{\sc iii}$]+$65&	9527	&FORS /VLT		& This paper					&	588	&23.8	& 0.81	\\
Y&	10212&	HAWK-I/VLT				& 	This paper				&	162	&24.9	& 0.52	\\
NB1.06&	10619	&HAWK-I/VLT			& 	This paper				&	150	&23.9	& 0.81	\\
J&	12511&	WFCAM/UKIRT			& 	H13			&	11190	&24.9 	& 0.79	\\
J&	12582&	HAWK-I/VLT				& 	This paper				&	104	&24.5	& 	0.60\\
H&	16383&	WFCAM/UKIRT			& 	H13			&	6000		&24.2 	& 0.84	\\
Ks&	21545&	HAWK-I/VLT				& 	This paper				&	74	& 23.8	& 0.43	\\
K&	22085&	WFCAM/UKIRT			& 	H13			&	12450	&24.6 	& 0.78	\\
IRAC1&	35573	&IRAC/{\it Spitzer}		&SpUDS; P.I. J.S. Dunlop&	--	&24.2 & 	1.7\\
IRAC2&	45049	&IRAC/{\it Spitzer}		&SpUDS; P.I. J.S. Dunlop&	--	&24.0 	& 1.7	\\
\hline
\end{tabular}
\caption{\label{table_data}Image data used to obtain photometry of \clustername}
\end{table*}

\subsubsection{Photometric catalogue}
The higher resolution HAWK-I $Y,\,J,\,Ks$ images were convolved to match the 0.81\arcsec\ PSF of the HAWK-I NB1.06 and FORS $[$S{\sc iii}$]+$65 images. To do this we identified 12 bright, unsaturated and isolated stars in the images and created growth curves in each of the bands with apertures between 0.2\arcsec\ and 8\arcsec. The HAWK-I $Y,\,J,\,Ks$ images were then convolved with a series of 16 Gaussians of $\sigma$ ranging from 1.5 to 3 pixels, and the growth curves remeasured in each of these smoothed images. For each band the Gaussian-smoothed image which resulted in the minimum-$\chi^2$ when compared to the HAWK-I NB1.06 growth curves was identified as the best PSF-matched image. We tested the PSF-matching algorithm by comparing the growth curves of stars in the convolved images. We found that the growth curves of the PSF-matched images for all bands are within 1.5\% for apertures of 2\arcsec -diameter or larger.

Fluxes were then measured in the HAWK-I and FORS images within 2\arcsec\ diameter circular apertures on each position of the $K-$selected catalogue of H13 using the {\sc idl} function {\sc aper}. Uncertainties were taken to be the square root of the photon counts in the apertures plus the standard deviation of the total photon counts within 2\arcsec-diameter apertures placed in regions with no object detections. Objects that fell within regions of the narrow-band images for which the exposure time was less than $30$\% of maximum exposure were removed from the catalogues to obtain data of approximately the same quality across the field of view.

To concatenate the new photometry with the H13 catalogue we first applied aperture corrections to the new photometry to account for the different PSFs. Fluxes were measured in 2\arcsec-diameter apertures in the new $J-$band HAWK-I image (smoothed to a PSF of 0.81\arcsec) and compared to the $J-$band fluxes from 2\arcsec-diameter apertures in the H13 catalogue. The $J-$band filters of the HAWK-I and WFCAM instruments have very similar response curves (see Fig.\,\ref{SED_examples}), so the flux of an object in both images should be the same.  A linear fit to the $J-$band fluxes showed that the new photometry was a factor of 1.06 times brighter than the $J-$band of the H13 catalogue due to the sharper PSF, hence an aperture correction of 0.94 was applied to the new photometry before it was concatenated with the H13 catalogue to form a complete photometric catalogue. All photometry was corrected for Galactic extinction using the maps of \citet{Schlegel1998}, and we rejected bright stars. The final catalogue consists of 3019 galaxies.

\subsubsection{Spectroscopic redshifts}
At the time the protocluster catalogue was created spectroscopic redshifts had been obtained for 62 galaxies in the 50 arcmin$^2$ area that is covered by the above images. These redshifts were taken from \citet{Papovich2010}, \citet{Tanaka2010}, and \citet{Santos2014}. In addition to these published redshifts, there exists a collection of unpublished redshifts from the UDSz (Almaini et al. in prep) and others ( \citealt{Simpson2012}; Akiyama et al.~in prep and \citealt{Smail2008}), which are collated and available through the UDS website\footnote{\rm http://www.nottingham.ac.uk/astronomy/UDS/data/data.html}. Sixteen of the 62 galaxies have spectroscopic redshifts that indicate they are part of the protocluster. 

After the protocluster galaxies were selected, two new catalogues of this region were published. \citet{Tran2015} provide spectra for 69 galaxies in the overlapping area, some of which had previous spectroscopic redshifts. 3D-HST is a slit-less grism spectroscopic survey which covers a ninth of the protocluster region that we imaged  \citep{3DHST}. We did not use these catalogues to select the protocluster members, instead we use them to estimate the completeness and contamination of the derived protocluster sample in Section\,\ref{sec:contamination}.

\subsection{Identifying protocluster and field galaxies}

\subsubsection{Photometric redshifts}
Photometric redshift probability distribution functions, $P(z)$, were determined for each source by fitting spectral energy distribution (SED) templates to the photometric data points using {\sc eazy}  \citep{Brammer2008}. We applied the apparent $K$-band magnitude prior that is distributed with the  {\sc eazy} package. In addition to the six templates supplied with the {\sc eazy} package, we also add the seventh template created by H13 by applying a small amount of Small Magellanic Cloud-like extinction to the bluest template from the {\sc eazy} set. Given the broader PSF and the effects of confusion the IRAC photometry is very uncertain, so photometric redshifts were determined both with and without the IRAC data points, and no significant difference was seen in the redshift probability distribution functions.

\begin{figure}
\includegraphics[width=1\columnwidth, angle=0]{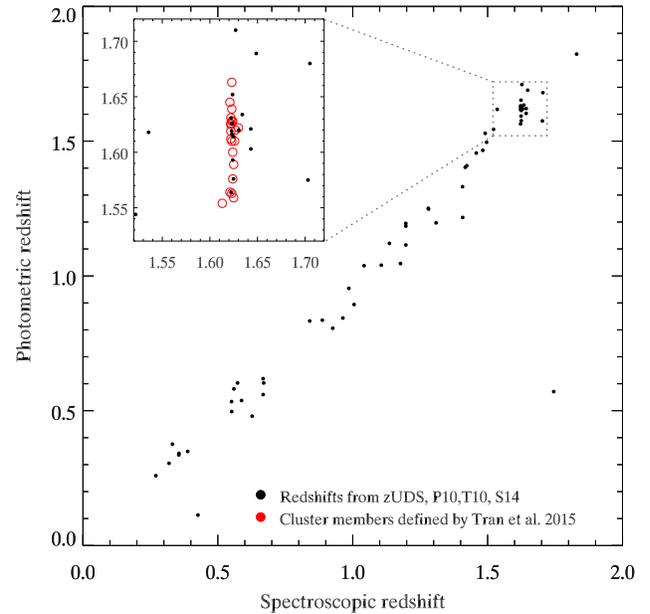}
\caption{\label{photoz_comparison}A comparison of the derived photometric redshifts with the spectroscopic redshifts for 62 galaxies (black points) from the original UDS spectroscopic sample. The red circles are spectroscopic redshifts from the targeted survey of \citet{Tran2015}. Photometric redshifts are the mean after prior redshifts (parameter z\_m2 outputted from the {\sc EAZY} photometric redshift fitting code).}
\end{figure}

Using the 62 spectroscopic redshifts as guides we made minor adjustments to the photometric zeropoints to ensure we obtained the most accurate photometric redshifts. We show a comparison of $z_{phot}$ versus $z_{spec}$ for all 62 spectroscopically confirmed galaxies in Fig.\,\ref{photoz_comparison}.  The biggest outlier at $z_{spec}=1.75$ is an X-ray source, hence it is likely that AGN emission is responsible for the poor template fit. 

The accuracy of the photometric redshifts at $z\sim1.6$ is twice as good compared to galaxies at other redshifts. The dispersion of $z_{phot}-z_{spec}$ for the 16 galaxies in the protocluster is $\Delta z/(1+z)=0.013$, which is half that of the full sample. This is because the FORS $[$S{\sc iii}$]+$65, HAWK-I 1.06$\micron$ and $Y-$band images sample the Balmer and 4000\AA\ breaks at multiple points for galaxies at the protocluster's redshift (see Fig.\,\ref{SED_examples}). This precision is similar to that achieved by the NEWFIRM Medium-Band Survey (NMBS; \citealt{Whitaker2011}) which augmented the optical and near-infrared photometry on the COSMOS and AEGIS fields with five medium band images that ranged across the wavelength range typically spanned by $Y$, $J$ and $H$. Here we show that a similar precision can be achieved for the galaxies in the protocluster with a smaller number of images because the filters were optimally chosen to span across the important Balmer and 4000\AA\ breaks.

\subsubsection{Selecting potential protocluster galaxies}

We use the full redshift probability distribution function to determine the likelihood of a particular galaxy being at the redshift of the protocluster. We determine the probability, \prob, of an object lying within a redshift range of $z\pm\delta z$ by integrating the redshift probability distribution functions,  $\int P(z)dz$, over the interval $z_{pc}+\delta z$ to $z_{pc}-\delta z$.  We first define two probabilities: (i) \prob$_{2\sigma}$ is the probability of the galaxy lying within $z_{pc}\pm0.068$ (i.e.\,within the 95\% confidence interval on the photometric redshifts), and (ii) \prob$_{5\sigma}=z_{pc}\pm0.17$. Then we use the spectroscopic redshifts to identify the best combination of these probabilities to select the potential protocluster members.

Selecting the {\it best} sample of protocluster galaxies depends on the purpose of the sample. For example, a clean sample of protocluster galaxies with few contaminants will likely miss many of the protocluster galaxies, but a complete sample of protocluster galaxies will likely contain a large number of contaminants. 

The compromise between a clean and complete protocluster galaxy catalogue is the \lq Goldilocks\rq\  sample, comprising of 143 galaxies that match the criterion  \prob$_{5\sigma}>90$\% and \prob$_{2\sigma}>50$\%. This Goldilocks sample was chosen to maximise the number of spectroscopically confirmed cluster members and minimise the number of interlopers. Fifteen out of sixteen of the spectroscopically confirmed cluster members are selected by this criterion, and only two spectroscopic interlopers (at $z=1.536$ and $z=1.703$) are included in the sample. We explore how the contamination and completeness of the selected protocluster candidates vary as a function of \prob$_{5\sigma}$ and \prob$_{2\sigma}$ in Section\,\ref{sec:contamination}. Additionally, not all of the Goldilocks sample will fall into the cluster by $z=0$, so this sample contains some non-protocluster members. The identification of these outliers is explored in Section\,\ref{Interlopers}.

Selecting protocluster members by integrating $P(z)$ means that our galaxy sample will be biased against objects that have a broad $P(z)$. Due to the multiple sampling of the Balmer and 4000\AA\ breaks the broadness of $P(z)$ is approximately the same for both blue and red galaxies of a given magnitude. However, the broadness of $P(z)$ is correlated with the signal to noise of the photometry. Since passive galaxies have lower mass-to-light ratios than star forming galaxies, this biases our protocluster galaxy selection against low-mass passive galaxies because they are faint.

\subsubsection{Selecting control field galaxies}
\label{sec:selectingPCGal}
The protocluster extends across the entire field of view covered by the narrowband data, therefore we cannot select a sample of field galaxies at the same redshift as the protocluster using the sample criteria as defined above. Instead we define a field sample which contains galaxies at redshifts slightly below and above the protocluster's redshift. 

The field sample is selected following the same criterion as the Goldilocks cluster sample, except that galaxies are selected around the redshifts $z_{field}=1.45$ and $1.81$, rather than centred on the cluster redshift. These redshifts are chosen as they are as close to the protocluster but avoid contamination by the protocluster galaxies themselves. The narrow and $Y$  bands lie nearby to the Balmer and 4000\AA\ breaks at these redshifts, and so help constrain the $P(z)$ distribution, but the dispersion is approximately twice the dispersion of galaxies at $z=1.62$. 

The effect of having a greater dispersion is that fewer control field galaxies are selected by the strict $\int P(z)dz$ criteria in comparison to the protocluster galaxies. Therefore the control field is likely to have a lower completeness than the protocluster field. The contamination level is likely to be similar as the interlopers in both samples have a similar $P(z)$ dispersion. 

The ideal control sample is selected from the same volume as the protocluster sample, and has similar levels of completeness and contamination, i.e.\,$89$\% and $12$\%, respectively (see Section\,\ref{sec:contamination}). We use 9 spectroscopic redshifts in the redshift range of the control field ($1.4<z<1.5$ and $1.76<z<1.86$) to estimate the completeness and contamination. Control field galaxies selected with \prob$_{5\sigma}>90$\% and \prob$_{2\sigma}>50$\% have a completeness of $67\pm27$\% with no contamination. However, it is highly likely the contamination is at least the same as in the protocluster sample, but we do not have sufficient spectroscopic redshifts to measure this accurately. Thus within uncertainties this criteria selects a control sample that has similar levels of completeness and contamination as the protocluster sample. It is therefore appropriate to use the same  $\int P(z)dz$ criteria to select the control sample. 

To obtain a high completeness of 88\% we must relax the selection criteria to \prob$_{5\sigma}>70$\% and \prob$_{2\sigma}>50$\%. The contamination of this sample is 43\%, which is far too high to robustly identify group structure, and even some spectroscopically confirmed protocluster members are selected in the \lq control field\rq\ sample. For these reasons it is not appropriate to select a control sample using such relaxed parameters. 

We therefore use the same  $\int P(z)dz$ criteria, which select a control sample with a similar level of contamination, and only slightly lower completeness. Comparing the different completeness rates we estimate that approximately 22\% of the galaxies are missing in the control field in comparison to the protocluster field. This missing fraction is highly speculative because we have so few spectra in the control field. However, we have examined each of our results in Sections \ref{results} and \ref{discussion} taking this into account and find that none of our conclusions are compromised if the completeness is reduced by 22\%. 

The control field sample contains 88 galaxies in total, 78 of which lie at $z\sim1.45$ and 10 lie at  $z\sim1.81$. The difference between the number of galaxies at the lower and higher redshift intervals is due to both cosmic variance, and cosmic dimming, which results in a $\Delta m=0.6$\,mag difference between galaxies at $z\sim1.45$ and 1.81. This field sample lies within a volume that is approximately twice the size of the Goldilocks protocluster sample.

The control field is selected from a comparatively small area and thus is subject to cosmic variance. At $z\sim1.45$ and $\sim1.81$ the control field is 17\% denser than the UDS as a whole. All of the enhanced density is due to a larger density of galaxies at $z\sim1.81$ in the small field of view. In the entire UDS there is an underdensity of galaxies at $z\sim1.8$ (H13), so the larger density in the small field of view used in this work may be more typical of the Universe. 

\subsection{Galaxy properties}
We derived stellar masses of the galaxies by fitting the $U$ to 4.5\micron\ photometry with stellar population models using the spectral energy distribution (SED) fitting code {\sc fast}  \citep{Kriek2009a}. The photometric catalogue was first scaled by the $K-$band ratio of flux measured using the {\sc sextractor} BEST aperture to that measured in a 2\arcsec\ aperture. The BEST flux is taken as the {\sc Sextractor} AUTO flux when no neighbour biases the results by more than 10\%, otherwise it is taken as the ISOCOR flux \citep{Bertin1996}. To fit the photometry we use \citet{BC03} stellar population templates with solar metallicity, exponentially declining star formation histories ($SFR=exp(-t/ \tau)$), the \citet{Kriek2013} average dust law (with $E_b= 1$ and $\delta=0.1$) and we assume stars are formed with the initial mass function of \citet{Chabrier2003}. Throughout the fitting process the redshifts of the protocluster members were fixed to $z=1.6233$ whilst the redshifts of the field galaxies were fixed to the mean after prior redshifts output from {\sc eazy}.

The observed star formation rates (SFRs) were measured from rest-frame UV luminosities using the \citet{Kennicutt1998} conversion from 2800\AA\ (assuming a Chabrier IMF): ${\rm SFR (}\Msun yr^{-1}{\rm )} = 8.24 \times 10^{-29} L_{2800} {\rm(}\ergpsphz \rm{)}$. To measure $L_{2800}$ we use the mean after prior redshifts (for field galaxies) or fixed $z=1.6233$ (for cluster members) to determine the filter that has the closest central wavelength to 2800\AA\ then add a $k-$correction based on linear interpolation taking into account the UV slope. In this work we are only interested in the distribution of galaxies of a certain observed SFR, so we make no correction for dust extinction.

\subsection{Identifying substructure}

\subsubsection{Measuring local environment} 
\citet{Muldrew2012} showed that the best measures of internal halo properties were nearest neighbour and Voronoi tessellation methods. Although the protocluster studied here does not consist of only one halo, the same principle applies as we wish to identify groups within a large high-density region.

We calculated the local environment of each location in the field of view using several methods: projected Voronoi tessellation; third, fifth and eighth projected nearest neighbour; cumulative distance from the first to fifth and tenth nearest neighbours. We adapted the original definition of the cumulative distance to the nth nearest neighbour \citep{Cowan2008} to be a measure of the projected environment:
\begin{equation}
\phi_{n\,{\rm th}}=\frac{1}{\rho\Sigma^{n}_{i=0}r_{i}^{2}}, 
\end{equation}
where $r_i$ is the projected distance to neighbour $i$ and $\rho$ is the total number of galaxies divided by the total area of the field-of-view. 
All of these methods are affected by edge effects, which means galaxies close to masked regions and the edge of the detector have unreliable environment measurements. We found there were strong correlations between all of these environmental measurements. 

The best measure was the cumulative projected $5$th neighbour distance, $\phi_{5\,{\rm th}}$, since this method improves the robustness of the nearest neighbour measures by minimising distortions by interlopers. Appendix B of \citet{Ivezic2005} describes in detail the improvement gained from using the distances to all $n$ nearest neighbours compared to only the distance to the $n$th neighbour. The choice of 5th nearest neighbour was regulated by survey depth; $n$ was chosen to be the largest possible number that is small enough so we can still identify structure on the scale of a few hundred kpc.

\subsubsection{Maps of galaxy and stellar mass density} 
\begin{figure*}
\includegraphics[height=1.04\columnwidth, angle=0]{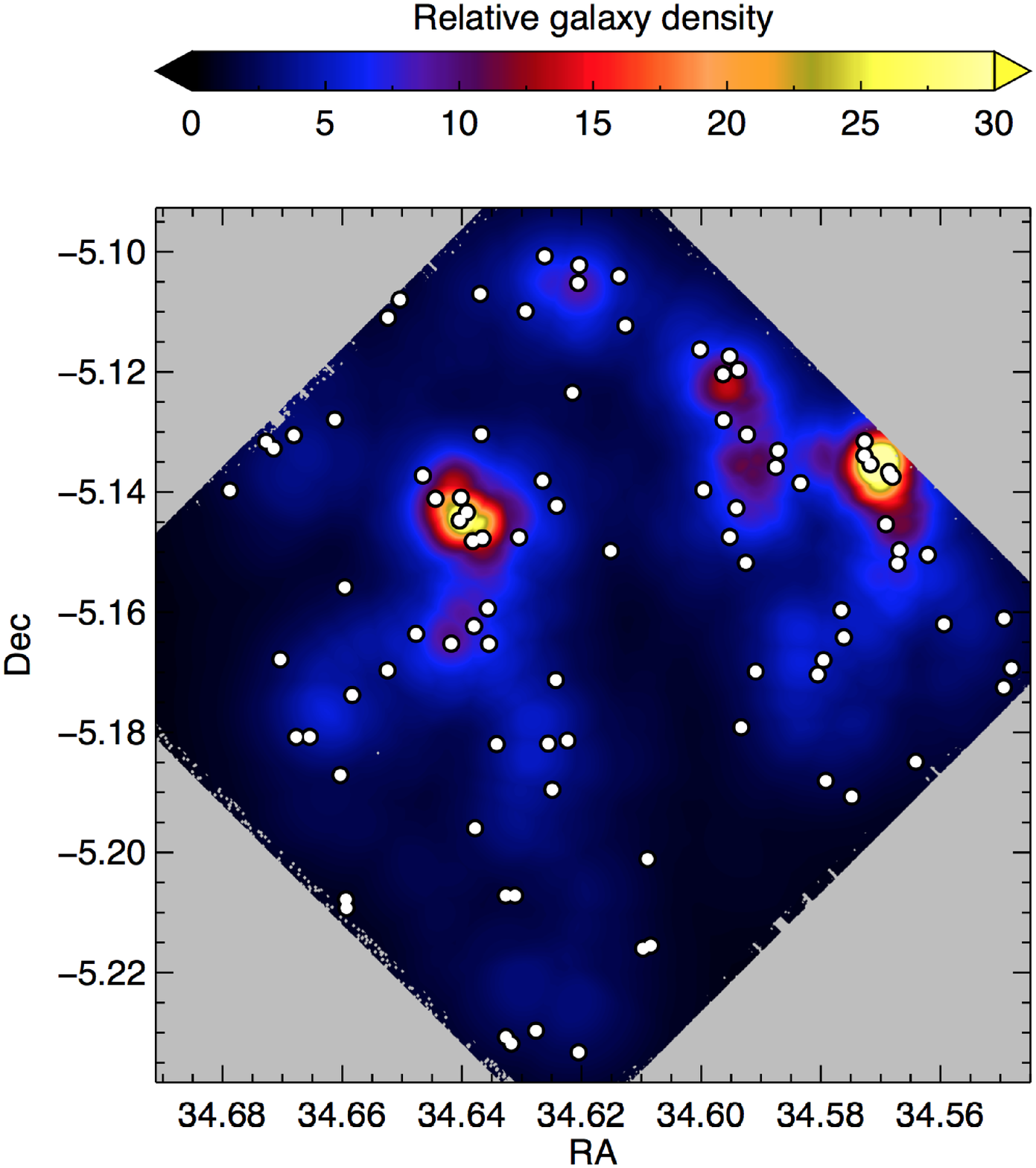}
\includegraphics[height=1.04\columnwidth, angle=0]{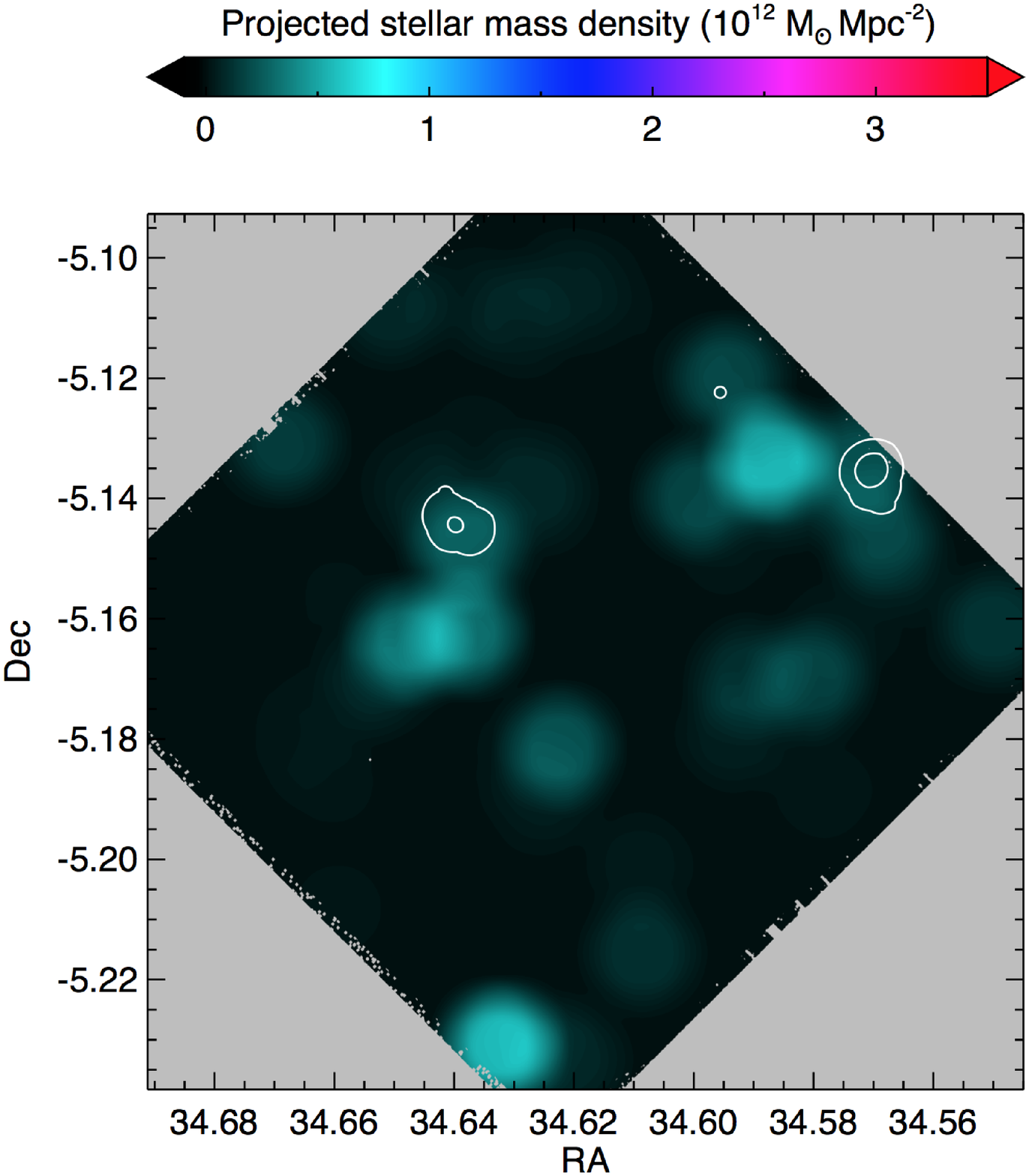}
\caption{\label{field_maps} Left: The 2D distribution of the galaxy density in the control field. The white circles mark the location of the galaxies and the background colour scale indicates the relative galaxy density determined by the cumulative distance to the 5th nearest neighbour.  Right: The projected stellar mass distribution in units of $10^{12}$\,\Msun~Mpc$^{-2}$.  The white contours mark the galaxy density as shown in the left panel.  The control field is twice the volume of the protocluster.}
\end{figure*}

Maps of the projected galaxy density of the protocluster and control field were created by measuring $\phi_{5\,{\rm th}}$ for each 0.25\arcsec\ pixel of $7.5\arcmin\times7.5$\arcmin\ field of view.  Maps of the projected stellar mass density were created by measuring the stellar mass density within 30\arcsec\ of each 0.25\arcsec\ pixel centre, and then smoothed with a boxcar average of 100 pixels (25\arcsec) width. Stellar mass density maps were scaled to display the stellar mass per Mpc$^{2}$ (Figs.\ref{field_maps}, \ref{density_maps} and \ref{density_galtype}). 

The galaxy formation models of \citet{Henriques2015} show us that stellar mass is a good tracer of the underlying dark matter, so the stellar mass maps give the most accurate representation of the distribution of dark matter in the protocluster. However, the resolution of the maps we can create is poor ($\sim20\arcsec \sim0.17$\,Mpc) and only very massive groups can be identified in these maps. The filaments that surround the groups of a protocluster may be narrow and will be missed in our stellar mass density maps. The galaxy density maps are more useful for locating the lower mass groups and to visualise the relatively low density filaments.  In addition to this complication, we must remember that the stellar mass of a halo does not linearly correlate with dark matter mass. At $z=1.62$, models predict that the stellar mass of a central galaxy increases approximately linearly with halo mass until the mass of the galaxy reaches $10^{10.5}$\Msun\  \citep[e.g][]{Wang2013}.  Galaxies with higher stellar masses have a higher total-to-stellar mass ratio. Thus a small group of low mass galaxies may trace a lower dark matter halo mass than a single massive galaxy of the same total stellar mass. 

In Fig.\,\ref{field_maps} we show the projected galaxy and stellar mass density maps of the control field to provide a comparison for the protocluster distribution shown in the following section. The control field comprises a region with twice the volume of the protocluster (a redshift interval both in front and behind the protocluster), so we have halved the projected stellar mass density to allow for a direct comparison with the protocluster.

The control field galaxy density map reveals two dense galaxy groups (see Fig.\,\ref{field_maps} and Table \,\ref{table_groups}). One of the groups lies very close to the edge of the field of view, which illustrates that using the $\phi_{5\,{\rm th}}$ allows us to identify galaxy groups up to 13\arcsec\ from the edge of the 7.5\arcmin\ field of view. The projected stellar mass density map shows a very different picture: the map is smooth which implies there is no significant structure in the projected dark matter distribution. The dense groups in the field therefore trace relatively low dark matter mass halos, or are line of sight projections of galaxies masquerading as groups. The regions of highest stellar mass density are not co-spatial with the location of the galaxy groups, but rather near regions that host the most massive galaxies that are not obviously in groups.

\subsubsection{Defining groups} 
Since the environmental parameter we define is a relative measurement that depends on the redshift and survey depth, we use the control field sample to define the value of  $\phi_{5{\rm th}}$  that best defines which galaxies belong to groups and which lie in between the groups. We visually inspect the $\phi_{5{\rm th}}$ maps and define separate groups as regions where the galaxy density peaks. Through trial-and-error we find the best divide of field and group galaxies for the control field sample occurs at  $\phi_{5{\rm th}}=13$. This boundary selects groups on the scale of a few hundred kpc, which is the scale of small collapsed galaxy groups expected at this redshift.

Approximately 90\% of the field galaxies have  $\phi_{5{\rm th}}<13$. The ten control field galaxies with $\phi_{5{\rm th}}>13$ lie within two highly concentrated groups of four and six galaxies within 30\,arcsec diameter apertures. We then apply the same  $\phi_{5{\rm th}}>13$ cut to define group galaxies within the protocluster sample. The difference in luminosity completeness due to the different redshifts does not matter because $\phi_{5\,{\rm th}}$ is defined as the relative density within a particular sample. The dividing line of overlapping groups is taken to be where  $\phi_{5{\rm th}}$ is a minimum between the groups.

The selection of galaxies defined as group members does not strongly depend on the choice of density estimator, or the choice of the number of nearest neighbours. This is because the relative density of each galaxy is very similar for $n$ between 3 and 10.  Since the boundary between group and intergroup galaxies is not absolute, but rather is defined as the density that selects the two groups in the control field, we select approximately the same galaxies as group galaxies regardless of the choice of $n$. We tested our results for $n$ ranging between $3$ and $8$ and found no significant difference in the number, size or total stellar masses of the groups identified in the protocluster. Therefore our results and conclusions are robust against changes in the choice of $n$. However, for $n> 10$ the spatial resolution of the density map is not sufficient to pick out groups on the scale of a few hundred kpc in either the control field or the protocluster.
 
\section{Results}
\label{results}
\subsection{Completeness and contamination level of the protocluster galaxy sample}
\label{sec:contamination}
We use the spectroscopic cluster sample of \citet{Tran2015}, and the grism sample of \citet{3DHST} to determine the completeness and contamination level of the protocluster sample. \citet{Tran2015} presents data on 109 good quality spectra of galaxies in and near \clustername. The targeted galaxies were bright and priority was given to galaxies with photometric redshifts 
 that lay close to the cluster's redshift. The images presented in this work only cover part of the region from which the spectra of \citet{Tran2015} were obtained, so only 69 spectroscopically observed galaxies are covered by our images.  All redshifts are determined through line emission so all spectroscopically measured objects host an active galactic nucleus or significant star formation.

\citet{Tran2015} defines {\it cluster} members as those which have a redshift within $1.612<z<1.635$. The cluster is defined as the main halo of the protocluster, whereas protocluster galaxies may lie tens of cMpc beyond the main protocluster halo and can have relative velocities up to 2000\kmps\ \citep{Contini2016}. We therefore define protocluster galaxies as those with redshifts in the interval $1.59<z<1.67$. There are 35 spectroscopically confirmed galaxies in this redshift interval and in the area we survey. We identify 31 of these 35 galaxies as protocluster candidates using our photometric redshift method, resulting in a completeness of $89\pm5$\%.

We selected 35 out of 69 spectroscopic galaxies in the \citet{Tran2015} sample as protocluster candidates using our method. One of these has an ambiguous spectroscopic redshift as it is defined as a cluster member with $z=1.634$ by \citet{Tanaka2010}, but defined as an interloper by \citet{Tran2015} (although no redshift is published). From the rest of the 34 galaxies, four have spectroscopic redshifts outside $1.59<z<1.67$ and are therefore contaminants. Therefore the contamination rate of our protocluster catalogue is only $12\pm5$\%. 

The galaxies with spectra span a similar range of stellar mass and $i'$ magnitude as the bulk of the photometric-redshift selected protocluster sample ($9.2< \log {\rm M}/\Msun <{10.7}$ and $23<i'<25.5$) so the contamination rate of 12\% is likely to be correct for most mass and luminosity bins in our protocluster catalogue. The completeness, however, is a strong function of galaxy luminosity and type (Hatch et al.\,in prep.). The spectroscopic success rate is strongly biased toward emission line sources, so the completion rate of $89\pm5$\% should be considered the percentage of star forming or active galaxies with $\log {\rm M}/\Msun > {9.7}$ that can be detected. 

There is a bias in comparing our catalogue to the \citet{Tran2015} sample because their sample was selected for spectroscopic followup based on photometric redshifts. Neither our sample nor the \citet{Tran2015} sample would locate galaxies whose SEDs result in erroneous photometric redshifts. So our completeness may be overestimated. We therefore compare our sample to the grism survey of 3D-HST \citep{3DHST}, who make no preselection based on photometric redshifts. Forty of our protocluster galaxy candidates have redshifts derived from a fit to both the photometric and grism data, of which only five do not have 68\% confidence intervals that span $z=1.6233$ and are thus interlopers. So the contamination rate is $13\pm5$\%. 

We can select protocluster candidates from the 3D-HST catalogues using the the maximum likelihood redshift parameter as those which satisfy $1.59<z\_max\_grism<1.67$. Using this method we find 24 protocluster candidates which are bright enough to be selected by our photometric redshift method and data. Twenty-one of these are selected as protocluster members. So the completeness is $88\pm7$\%. It is reassuring that the completeness and contamination rates derived from two different data sets are in perfect agreement. 

\begin{figure}
\includegraphics[height=1.0\columnwidth, angle=-90]{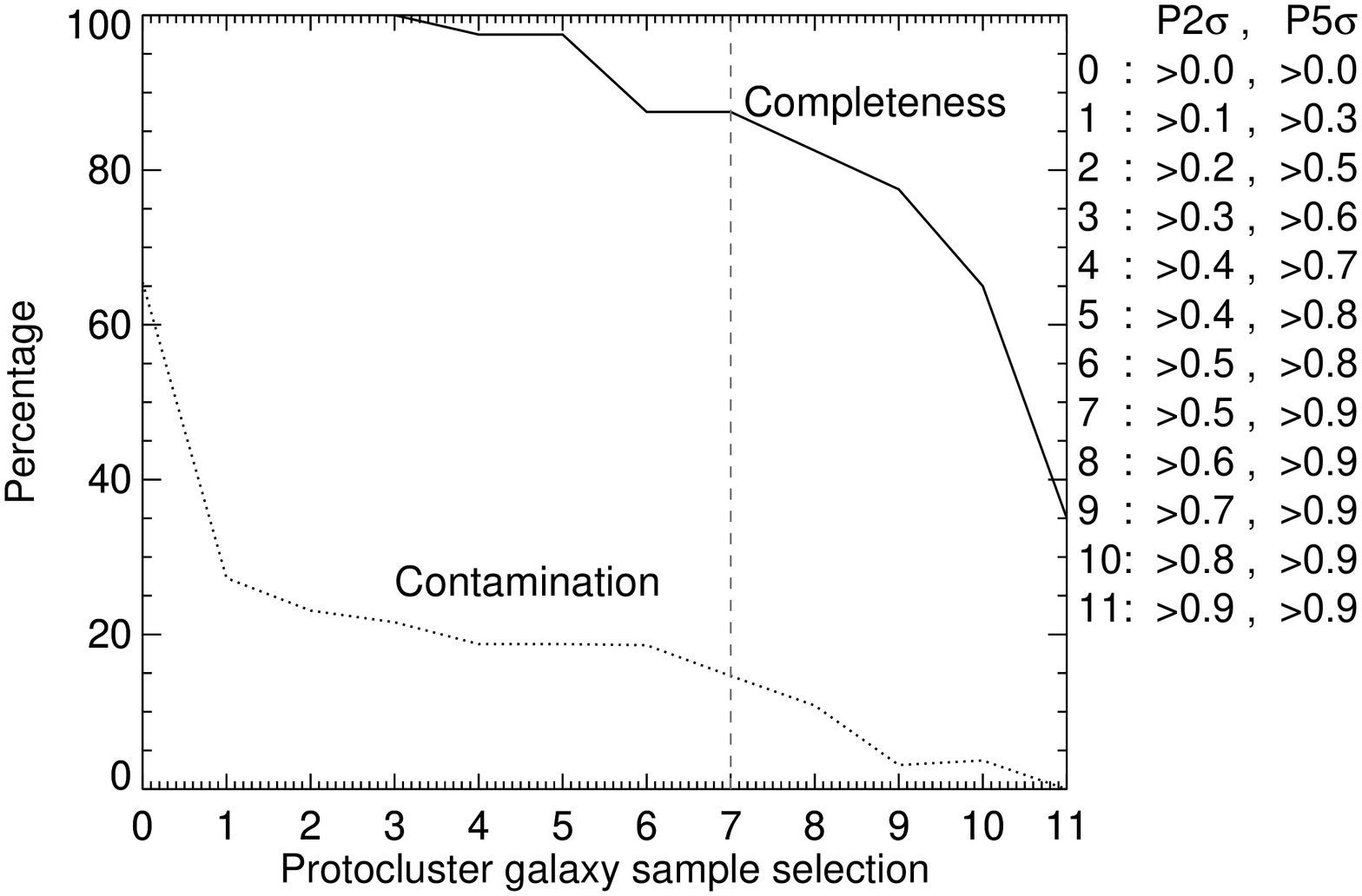}
\caption{\label{fig:selection}  The completeness and contamination of the selected protocluster galaxy sample for various  \prob$_{5\sigma}$\ and \prob$_{2\sigma}$ selection parameters. The optimal values for identifying protocluster structure occurs when contamination is $\lesssim15$\% and completeness is $\gtsim80$\%. This occurs when $0.8<$\prob$_{5\sigma}<0.9$ and $0.5<$\prob$_{2\sigma}<0.6$. The \lq Golilocks\rq\ sample (marked by the dashed vertical line) lies in the middle of this range.}
\end{figure}

In Fig.\,\ref{fig:selection} we use all spectroscopically confirmed galaxies to show how the completeness and contamination changes as the protocluster galaxy selection parameters (\prob$_{5\sigma}$\ and \prob$_{2\sigma}$) are varied. Neither the completeness or contamination vary greatly when the parameters range from  \prob$_{5\sigma}>0.8$ and \prob$_{2\sigma}>0.5$ to \prob$_{5\sigma}>0.9$ and \prob$_{2\sigma}>0.6$. Within this range, the number of protocluster galaxies selected varies only by $\pm25$, and all of the following results and conclusions are robust to this small change in the protocluster galaxy sample.

By choosing stricter or more relaxed selection criteria one can select either a cleaner sample or more complete sample of protocluster galaxies. Both of these samples are not appropriate for mapping the protocluster structure. A large number of interlopers smooth out the protocluster structure, so the groups are no long obvious when the selection parameters are less stringent than \prob$_{5\sigma}>0.8$ and \prob$_{2\sigma}>$0.4. On the other hand, there are too few galaxies to identify groups when the parameters are more stringent than \prob$_{5\sigma}>0.9$ and \prob$_{2\sigma}>0.7$. The optimal values for identifying protocluster structure occurs when the interloper fraction is less than approximately 15\% and the completeness (for bright star forming galaxies) is more than 80\%. For our data this occurs when $0.8<$\prob$_{5\sigma}<0.9$ and $0.5<$\prob$_{2\sigma}<0.6$.

\subsection{The structure of the \clustername\ protocluster}

\begin{table}
\begin{tabular}{|l|c|c|c|c|c|}
\hline
Group name&RA&  Dec & \#& Total stellar mass\\
&&  &  & (log$_{10}$\Msun)\\
 \hline
Group 1 & 34.5898 & $-$5.17217& 16 &  11.93 \\
Group 2&34.6194	&$-$5.20089	&  6	& 11.53	\\
Group 3&34.5734	&$-$5.16781	&  5 	& 11.39	\\
Group 4&34.5823	&$-$5.16906	& 6	&11.01	\\
Group 5&34.5980	&$-$5.15953	& 7	&10.96	\\
Group 6&34.6115	&$-$5.11375	& 7	&10.55	\\
\hline
Field group A   &34.57273	&$-$5.13392	&	6&  10.96\\
Field group B	&34.63656	&$-$5.14772	&	4&  10.67\\
\hline
\end{tabular}
\caption{\label{table_groups} A list of the groups within the \clustername\ protocluster and control field. Columns 2 and 3 list the RA and Dec of the group. Columns 4 and 5 lists the number and total ${\rm log}_{10}$ stellar mass of galaxies in each group. The field groups are similar to groups 4, 5 and 6 in the protocluster which are compact and have relatively low masses.}
\end{table}

\begin{figure*}
\includegraphics[height=1.04\columnwidth, angle=0]{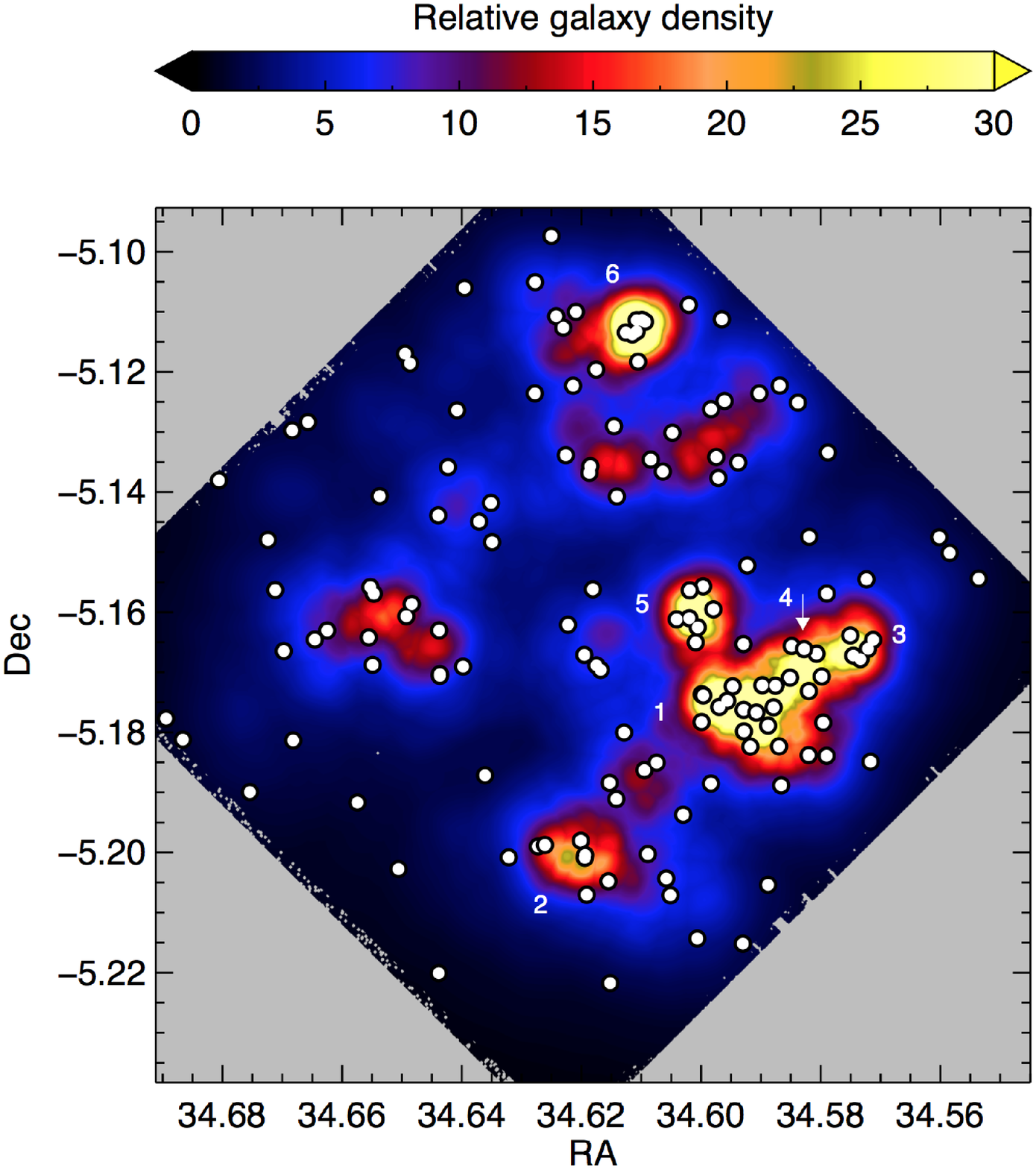}
\includegraphics[height=1.04\columnwidth, angle=0]{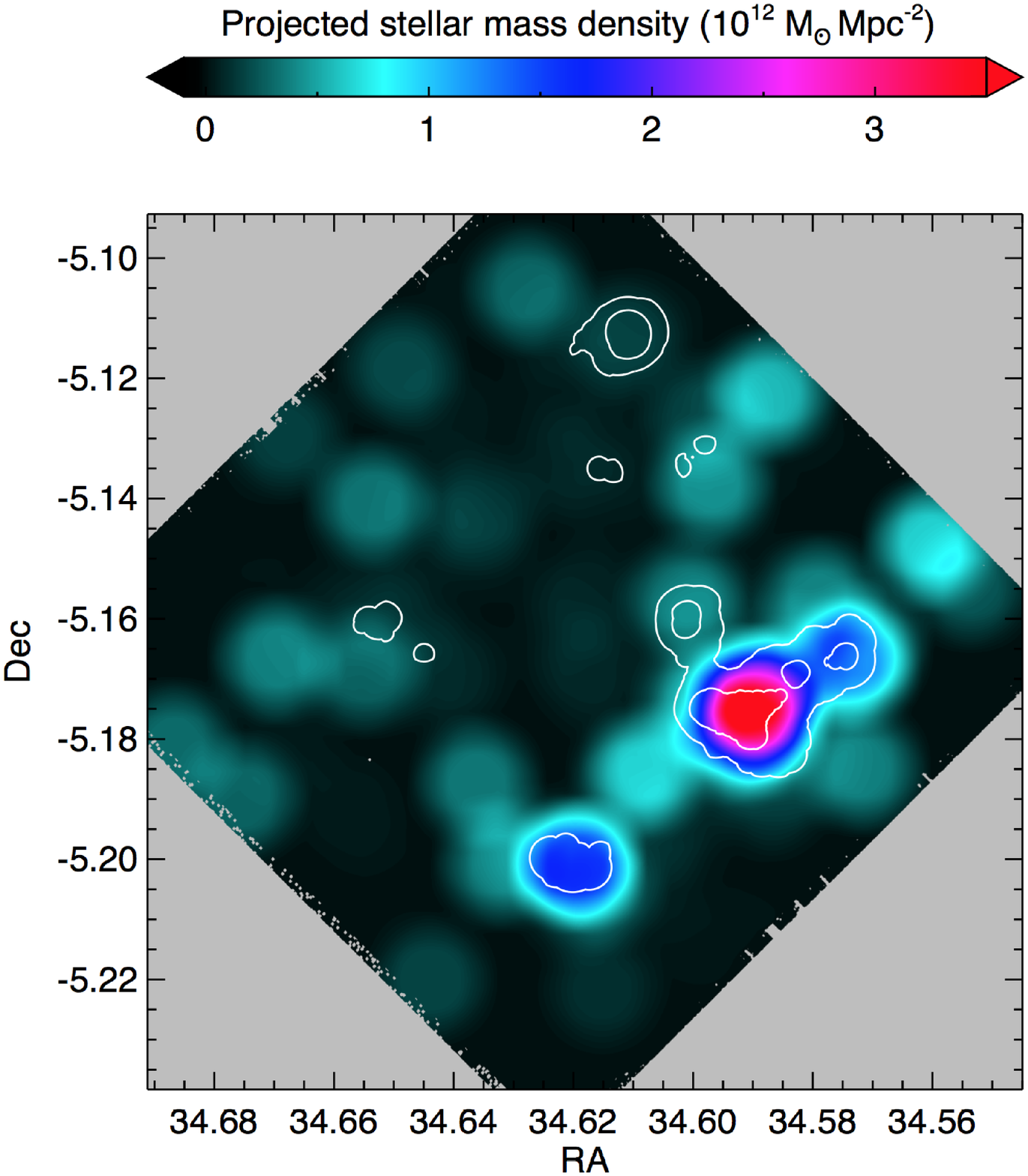}
\caption{\label{density_maps}  Left: The distribution of galaxies within \clustername. The background colour indicates the relative galaxy density. The protocluster structure consists of 6 galaxy groups, the largest of which is the nascent cluster (group 1), surrounded by a dense sea of intergroup galaxies. The properties of groups $1-6$ are given in Table \ref{table_groups}. Right: The distribution of stellar mass within the protocluster. The colour indicates the projected stellar mass density in units of $10^{12}$\Msun Mpc$^{-2}$. The white contours mark the galaxy density as shown in the left panel. The protocluster is dominated by the largest group and the two flanking high-mass groups (2 and 3). The total stellar mass in the groups (5 and 6) is relatively low and does not appear to be denser than the typical intergroup density within the protocluster. }
\end{figure*}

The structure of \clustername\ is displayed in Fig.\,\ref{density_maps}.  The structure of this protocluster differs from the control field in two distinct ways. First, the protocluster contains a greater number of galaxy groups, which generally contain much more stellar mass than the control field groups. Second, the galaxy density between the groups is much higher. Both of these structural features signify the presence of the protocluster.

\begin{figure*}
\includegraphics[height=1.04\columnwidth, angle=0]{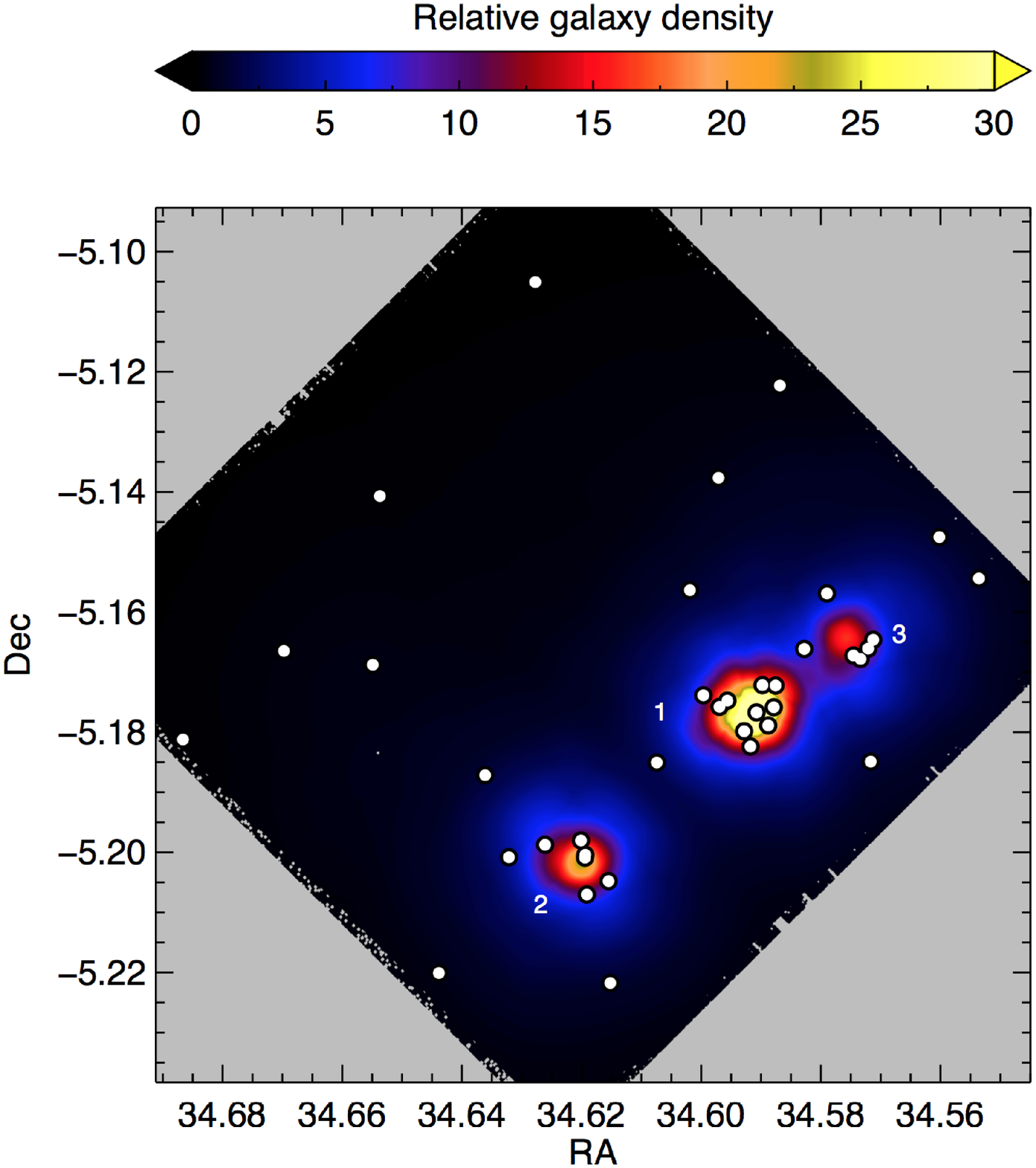}
\includegraphics[height=1.03\columnwidth, angle=0]{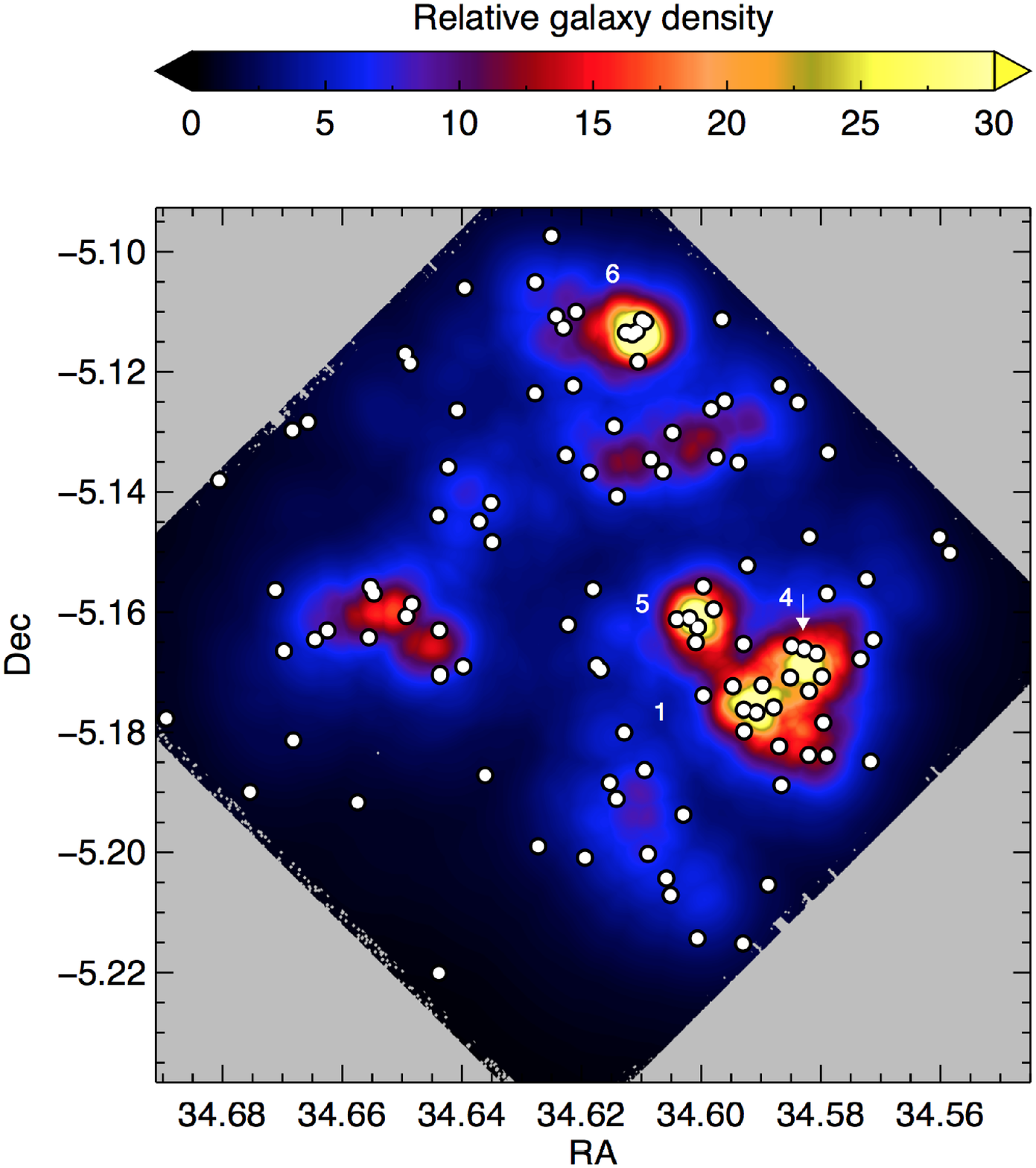}
\caption{\label{density_galtype} The galaxy density of red (left) protocluster galaxies with $z-J>1.3$, and galaxies with directly observed  SFR$ > 5$\Msunpyr\ (right). The background colour scale indicates the $\phi_{5\,{\rm th}}$ density measurement and the white circles pinpoint the location of protocluster galaxies with red colours or high SFRs. }
\end{figure*}
  
\subsubsection{Protocluster galaxy groups}
A key feature of the structure of this protocluster is the presence of several galaxy groups that are significantly more massive than those found in the control field. The galaxy density of each of the groups is not much greater than the control groups, but the stellar mass density is significantly larger, and they appear prominent in maps of projected stellar mass density.  The galaxies in the protocluster groups are, on average, more massive than those in control field groups.

The largest and most massive group (group 1) lies at the location of the tentative $4.5\sigma$ detection of X-ray emission reported by \citet{Tanaka2010}. The total mass of this group is estimated from the X-ray luminosity to be $5.7[\pm1.4] \times 10^{13}$\Msun, which is consistent with the mass estimated from the velocity dispersion of $\sigma_{group~1} = 254 \pm 50$\kmps\ \citep{Tran2015}. This dense core is surrounded by five additional galaxy groups, which are listed in Table 2. The separate groups can be more easily identified in Fig.\,\ref{density_galtype}. Group 2 also has a weak $1.5\sigma$ X-ray detection reported by \citet{Tanaka2010}. 

Groups 5 and 6 are compact groups with very high galaxy densities, but they contain relatively little stellar mass and are similar to groups A and B found in the control field. As this type of low mass group is found in both environments it is unlikely that these structures are unique signatures of the protocluster.  By contrast, groups $1-4$ are very different types of structures to the low mass groups in the control field. These groups contain more stellar mass than is found in either of the control field groups, and they are prominent features in the stellar mass density maps. Since the ratio of stellar to dark matter mass is expected to change once galaxies exceed $10^{10.5}$\Msun, it is likely that the dark matter mass distribution in this protocluster is even more skewed towards these four groups than the stellar mass density map suggests.

Many algorithms for finding protoclusters concentrate on locating large galaxy overdensities. Such overdensities are prone to detrimental line of sight projection effects and so protocluster candidate catalogues are plagued with contaminants. The maps presented here suggest that protoclusters may be more prominent as stellar mass overdensities rather than galaxy overdensities.

\subsubsection{Protocluster intergroup galaxies}
A distinct feature of the protocluster is the high galaxy and stellar mass density in between the groups.  We refer to these galaxies as the \lq intergroup\rq\ galaxies. The projected density of intergroup galaxies is 2.5 times greater than in the control field\footnote{If the control field sample is 22\% less complete than the protocluster (see discussion in Section \ref{sec:selectingPCGal}) then the projected intergroup density in the protocluster is 2 times that of the control field.}. Unlike the overdensity in the groups, there is no enhancement of the stellar mass density over the galaxy density: the projected stellar mass density of the protocluster as a whole is only a factor of two greater than the control field.  These results are robust against changes to the choice of the density estimator, or the value of $nth$ nearest neighbours. The enhancement in galaxy density does not lie immediately outside the groups, but rather is evenly spread across all of the intergroup region. 

Most of the protocluster galaxies lie between the groups. Within the limited field-of-view 11\% of the protocluster galaxies are in the most massive group, 22\% in the additional groups and the remaining 67\% reside between the groups. The fraction of galaxies in the most massive group is an upper limit as there are likely to be additional protocluster galaxies beyond the observed window.  Protoclusters extend over several Mpc so the observations presented here are not likely to encompass the entire protocluster structure \citep{Muldrew2015}.

\begin{figure*}
\includegraphics[height=2.13\columnwidth, angle=-90]{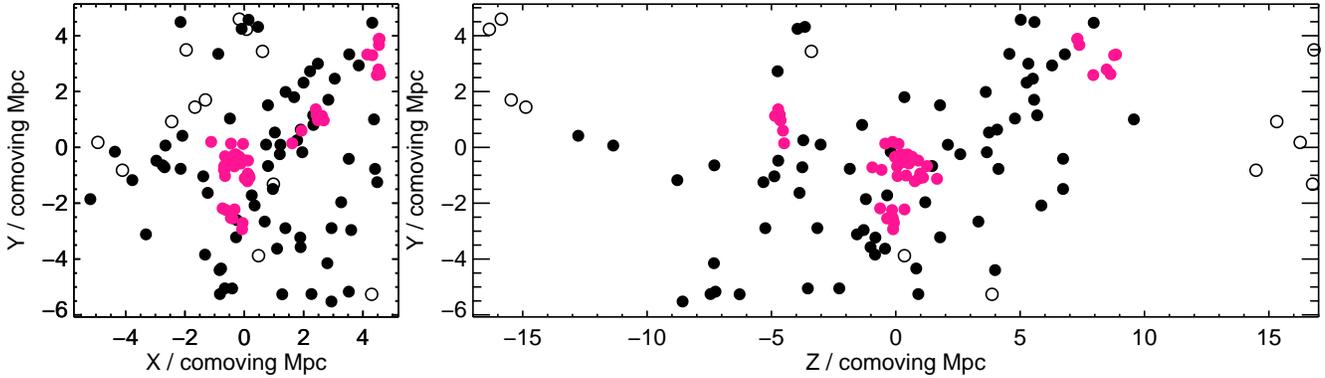}
\caption{\label{xysim} Distribution of $10^{9}$\Msun\ galaxies within the $10.2\times10.2\times34.0$\,cMpc box around one of the 40 Millennium Simulation protoclusters that have similar structural properties to \clustername.  The left plot shows the distribution in the x-y plane, whilst the right plot shows the z-y plane distribution. Group galaxies are pink whilst intergroup galaxies are black. Filled circles are galaxies that will merge to form the cluster by $z=0$ and the open circles are those that do not fall in by the present day and are therefore contaminants to the protocluster sample at $z\sim1.6$.}
\end{figure*}

\subsection{Biased views of protoclusters from observing only galaxy subsets}

Protoclusters are commonly observed by selecting only one type of member galaxy such as line-emitting galaxies \citep{Hayashi2012,Koyama2013}  or red galaxies \citep{Kajisawa2006,Hatch2011a}. In Fig.\,\ref{density_galtype} we explore what the structure of  \clustername\ looks like when we limit our observations to only red protocluster members ($z-J>1.3$), or star-forming members with an observed SFR $>5\Msunpyr$. The red population includes passive galaxies and star forming galaxies with significant amounts of dust obscuration ($A_V>1$\,mag). The star forming population is akin to unobscured populations such as Lyman break galaxies, and Ly$\alpha$, [O{\sc ii}] or H$\alpha$ emitters.  The structures of the protocluster revealed by the two types of galaxies are disparate. 

The most massive group appears in both maps, and both types of galaxies identify some of the other galaxy groups. The red galaxies locate the massive groups (1, 2 and 3), but do not find the lower-mass groups 4, 5 and 6. The converse is true for the star forming galaxies, which locate groups 1, 4, 5 and 6 but do not identify the massive groups 2 and 3. It is therefore possible that emission line maps of protoclusters are unable to locate some of the most massive groups in the protocluster. Red galaxies do a poor job of tracing the intergroup galaxies, whereas the structure of the intergroup filaments is well-traced by star forming protocluster galaxies.

Comparing the maps of Fig.\,\ref{density_maps} to Fig.\,\ref{density_galtype} illustrates that studying the protocluster with only one type of galaxy can severely bias our view of the protocluster. It is likely that galaxy groups would be entirely missed by studying the protocluster in either red or star forming galaxies, and the red galaxies do a poor job at locating the enhanced density of intergroup galaxies.  Protoclusters are therefore far more complex structures than their lower-redshift cluster descendants, whose structure can be traced well by galaxies that lie on a red sequence in colour-magnitude space. Galaxies are still rapidly forming in protoclusters and galaxy formation appears to be unevenly distributed.  To view the complete structure and see the full complexity of the protocluster we must identify all types of protocluster galaxies.

\section{Discussion}

\label{discussion}

Comparing observations of protoclusters and clusters at different redshifts can reveal how clusters form and galaxies evolve, but only if we can statistically link protocluster ancestors and cluster descendants that follow the same evolutionary paths. Clusters in the local Universe come in a variety of sizes, so to trace cluster formation it is imperative that we are able to distinguish the progenitors of different types of clusters in their protocluster state. 

The crux of the problem is determining what the end product of a protocluster will be and how it will evolve to get there.   Clusters form by the accumulation of galaxies from the field and by merging with smaller groups. This process is stochastic so the most massive clusters at any observed redshift will not necessarily become the most massive clusters by the present day (although they are statistically more likely to do so). By observing a large fraction of the \clustername\ protocluster we have determined what material is available to grow the cluster, and where that material is located. The structure of the protocluster gives us additional information to constrain the future evolution of the forming cluster. In this section we show how large N-body cosmological simulations can be used to determine a protocluster's evolution and present-day mass from observations of the protocluster structure.    

\subsection{Millennium Simulation counterparts to \clustername}

\label{sims}
We construct a sample of simulated galaxy protoclusters that have similar structural properties to \clustername\  using the \citet{Henriques2015} semi-analytic model applied to the Millennium Simulation \citep{Springel2005} scaled to the Planck Cosmology \citep{Planckcosmology2014}. The simulated box is periodic of side length 480 Mpc\,$h^{-1}$ where $h=0.673$. The closest snapshot to the observations of \clustername\ was $z=1.613$. We first select main halos with the same mass as the X-ray-derived mass of the main halo of \clustername\ by identifing 652 dark matter halos, in this snapshot, with masses in the range $4.3<{\rm M}/10^{13}\Msun<7.1$ that were the main halos of their cluster formation tree.  We then selected a $10.2\times10.2\times34.0$\,cMpc box around these halos and identified 40 \clustername--like protoclusters as the subset whose dark matter halos within this box have mass ratios similar to the stellar mass ratio of the groups in \clustername, i.e.~1 : 0.40 : 0.29 : 0.12 : 0.11: 0.04, with a range of $\pm0.1$ in each of these ratios. We chose the  $10.2\times10.2\times34.0$\,cMpc box size because the observed field of view is 10.2\,cMpc at the protocluster's redshift, and we chose a 34\,cMpc depth because the simulated galaxy overdensity in this volume is similar to the observed galaxy overdensity of \clustername.  It is reasonable to use the ratio of the stellar masses as a proxy for the total mass ratios since the total stellar mass of galaxy groups is proportional to the viral mass in the model of \citet{Henriques2015}.

Fig.\,\ref{xysim} shows the galaxy distribution of one of the 40 simulated protoclusters with a similar structure to \clustername. Similar to the real 2D distribution of \clustername, the protocluster consists of a small number of groups enveloped by intergroup galaxies which appear to have a random distribution across the whole field. Viewing the intergroup galaxies in the Z-Y plane allows us to see the wider structure. Large numbers of galaxies surround the groups and the protocluster core is characterised by a high density of these galaxies.

Protoclusters in simulations are large structures, with the most massive having galaxies spread across 50 cMpc ($\sim 32$\arcmin) at $z\sim2$. However, the groups are very centrally concentrated with $77$\% of groups residing within $\pm5$\,cMpc of the most massive halo. The intergroup galaxies are still centrally concentrated, but less so than the groups, with $52$\% of intergroup galaxies residing within the central $10$\,cMpc. To determine the structure of the protocluster it is important to observe all of the galaxy groups that make up the protocluster, whereas the intergroup galaxies are of less importance. So to measure the protocluster structure it is sufficient to obtain precise photometric redshifts over a small field of view of approximately $5-10$\arcmin\ radius centred on the most massive protocluster group. 

Throughout the rest of this article we refer to \lq\clustername-like protoclusters\rq\ and \lq protoclusters with a similar structure to \clustername\rq\ as protoclusters in the simulations which contain dark matter halos with the ratio 1 : 0.40 : 0.29 : 0.12 : 0.11: 0.04 within a $10.2\times10.2\times34.0$\,cMpc box.

In section \ref{sec:selectingPCGal} we discussed the possibility that the control field sample may be less complete than the protocluster by $\sim22$\% due to the higher photometric redshift errors at redshifts away from $z=1.62$. If the control field has an additional incompleteness to this level the depth of the protocluster increases from 34\,cMpc to 41\,cMpc. This does not affect the following analysis since our selection of \clustername-like protoclusters depends only on the galaxy groups, and most of the groups lie within the central $\pm5$cMpc of the protocluster. We obtain almost identical results in the following discussion whether the protocluster depth is 41.0\,cMpc or 34.0\,cMpc.

Observations only provide estimates of galaxy properties, such as stellar mass. Therefore in any observational survey the observed stellar mass distribution may be systematically biased, or have a wider distribution than the simulated stellar mass distribution. Our method for identifying \clustername-like protoclusters removes much of this bias by using the ratio of total stellar masses of the groups to determine the ratio of dark matter halo masses. Our method therefore crucially relies on the assumption that stellar mass directly traces dark matter, and that the total mass of group 1 is well constrained, but our method is not affected by the bias between observed and simulated galaxy properties.

\subsection{The $z=0$ mass of \clustername}
\label{sec:mass}

\begin{figure}
\includegraphics[height=1.\columnwidth, angle=-90]{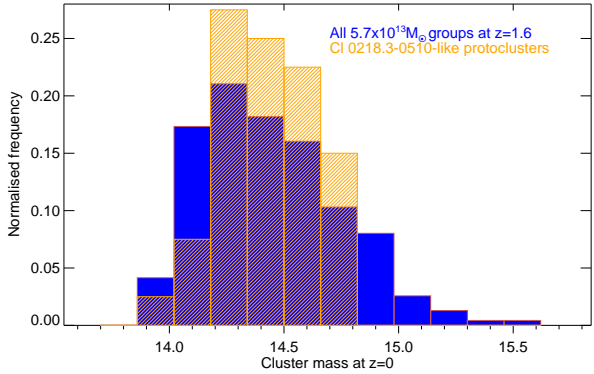}
\caption{\label{massz0}The blue histogram shows the range of  $z=0$ cluster masses from all 652 main halos with $M > 10^{13.6}$\Msun\ at $z=1.61$. The orange hatched histogram shows the range of present-day masses from the 40 simulated clusters with the same group distribution as in \clustername. }
\end{figure}

By mapping the structure of the protocluster we have constrained the allowed growth rate of the cluster and therefore limited the allowed range of present-day cluster mass ($M_{z=0}$). In Fig.\,\ref{massz0} we show the distribution of $M_{z=0}$ for all 652 dark matter main halos in the Millennium Simulation at $z=1.61$ with masses in the range $4.3<{\rm M}/10^{13}\Msun<7.1$ in blue, and those 40 protoclusters with the same structure as \clustername\ in orange. Without taking into account the protocluster structure $M_{z=0}$ is poorly constrained and the present-day cluster mass may be anything in the range of $10^{13.9-15.5}\Msun$. 

The 40 simulated \clustername-like protoclusters have a narrower range of $M_{z=0}$. All of the \clustername-like protoclusters become clusters with masses in the range $1.0<{\rm M}/10^{14}\Msun<6.6$. The median present day mass of \clustername-like descendants is  $2.7\times 10^{14}\Msun$, therefore it is likely that  \clustername\ will become a cluster with a slightly lower mass than the Virgo cluster ($4.4<{\rm M}/10^{14}\Msun<7.4$; \citealt{Hoffman1980}) by the present day.
 
This result relies on a number of assumptions: (i) that the initial mass of \clustername\ is $4.3<{\rm M}/10^{13}\Msun<7.1$; (ii) that our method unambiguously locates galaxy groups; and (iii) that stellar mass exactly traces dark matter mass in the groups. Of these three assumptions, the present day mass of \clustername\ most critically depends on whether we know the mass of the main halo at $z=1.62$. Independent analyses of X-ray data from both {\it Chandra} and {\it XMM-Newton} \citep{Tanaka2010, Finoguenov2010, Pierre2012} and dynamical velocity dispersion estimates \citep{Tran2015} are all consistent with our assumed initial mass, however we caution that estimates for group and cluster masses at this redshift are highly uncertain.

\subsection{Tracing ancestor protoclusters and descendant clusters}

\label{sec:evolution}

\begin{figure}
\includegraphics[height=1.\columnwidth, angle=-90]{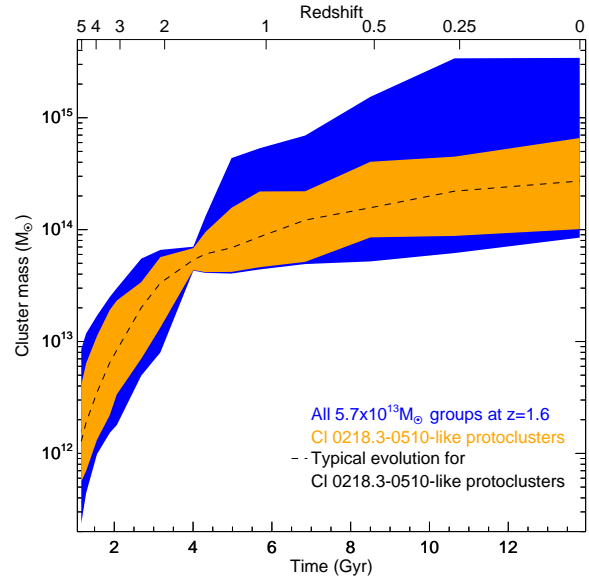}
\caption{\label{evolution}The range of evolutionary paths taken by \clustername-like protoclusters (orange), and all dark matter main halos with masses in the range of $4.3<{\rm M}/10^{13}\Msun<7.1$ (blue). The shaded region encompasses 99\% of all possible evolutionary sequences. The structure of the \clustername\ protocluster allows us to estimate the future growth of the cluster.}
\end{figure}

Mapping the structure of the protocluster at $z\sim1.6$ allows us to estimate its state at all redshifts. In Fig.\,\ref{evolution} we show the evolution of the 652 main dark matter halos in the Millennium Simulation with  $4.3<{\rm M}/10^{13}\Msun<7.1$ in blue, and for the \clustername-like simulated protoclusters in orange. Based on this, the most rapid period of growth for \clustername\ occurs at $z>2$. In 2.5 billion years the main halo of the protocluster grows by a factor of $20-100$. Its growth from the observed epoch of $z=1.6$ to the present is more muted, with only a factor of $2-8$ increase in mass. The clusters that grow more at $z<1$ than \clustername\ are either surrounded by more massive halos or have a larger number of nearby halos. Approximately 60\% of the faster growing clusters have more massive second-ranked halos, and 50\% have more mass in their six most massive halos compared to \clustername.

The structure of \clustername\ allows us to improve our estimates for its descendant mass at all redshifts, in particular it strongly constrains the upper limit of its mass at all redshifts. However very little improvement is made in constraining the mass of its ancestor protoclusters. The future growth of \clustername\ is constrained because we are able to estimate how much material is available for future consumption. But the wide-field protocluster structure does not relay any information about its main halo mass prior to the epoch of observation, so we are unable to constrain the ancestor protoclusters that will form \clustername. 

These results demonstrate that the structure of a protocluster can help constrain its evolutionary path. By mapping the structures of a large sample of protoclusters and clusters across $5>z>0$ we can place the (proto)clusters in evolutionary sequences that describe how clusters form. If we can reliably determine which galaxies in each observation will become cluster members, then the sequences of evolving clusters also provide samples of galaxy ancestors and descendants for a closed system of galaxies. Such sequences are powerful tools for studying the evolution of galaxies. 

\subsection{Is \clustername\ a typical ancestor of Virgo-mass clusters?}
\begin{figure}
\includegraphics[height=1.\columnwidth, angle=-90]{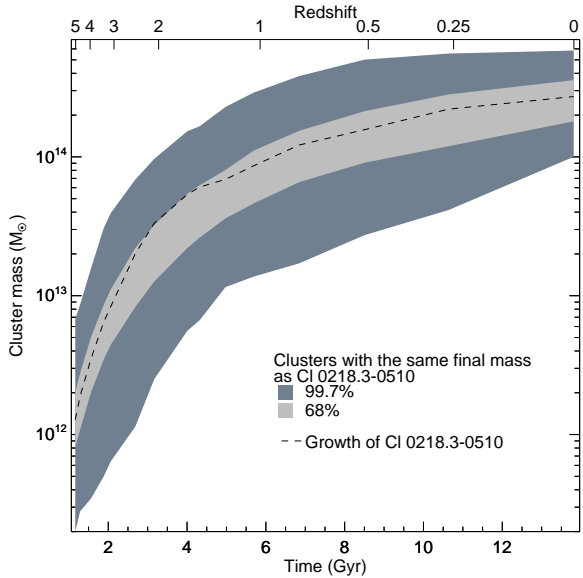}
\caption{\label{evo_back}A comparison of the evolutionary growth of \clustername\ (dashed line) with that of typical Virgo-like clusters that have final masses in the range  $1<{\rm M}/10^{14}\Msun<6$ (dark grey region). The light grey region marks the evolutionary growth of the most typical 68\% of protoclusters.  \clustername\ is typical of Virgo-like clusters, but contains slightly more mass in its main halo than is usual.}
\end{figure}

In Section \ref{sec:mass} we found that \clustername\ is likely to become a cluster of similar mass to Virgo by the present day. We can therefore ask the question, \lq do the progenitors of Virgo-like clusters all look like \clustername?\rq\ To answer this question we extracted from the Planck-scaled Millennium catalogue all dark matter haloes at $z=0$ with virial masses in the range $1<{\rm M}/10^{14}\Msun<6$. We selected from this a subsample that matched the $z=0$ mass distribution of \clustername\ as shown in Fig.\,\ref{massz0} and constructed their evolutionary paths, which we display in Fig.\,\ref{evo_back}. 

The evolutionary growth of \clustername\ is not atypical of Virgo-like clusters: its most likely evolutionary path is consistent with many clusters that end up with similar $z=0$ masses. However, the main group at $z=1.62$ is larger than average, with 77\% of clusters having lower masses at this redshift. This means that a larger fraction of the galaxies in \clustername\ will spend a longer time in the dense group environment than is typical for such clusters, and environmental quenching of star formation would have started early for a larger fraction of its members. A signature of this early assembly may be visible in the stellar populations of its descendant clusters as we would expect the mean stellar age of its member galaxies to be older than most clusters of similar mass.

\subsection{Which of the \clustername\ galaxies will become cluster galaxies}
\label{Interlopers}

\begin{figure}
\includegraphics[height=1.\columnwidth, angle=-90]{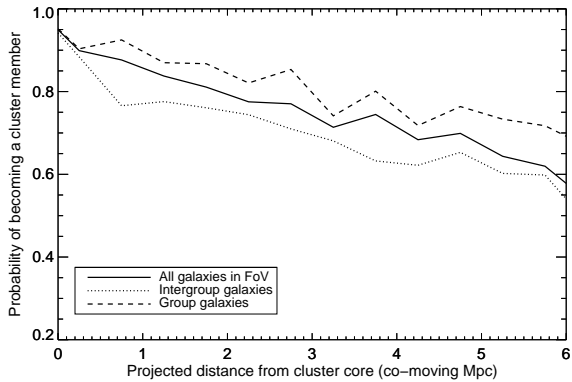}
\caption{\label{interloper_prob} The probability of galaxies within \clustername-like protoclusters in the Millennium Simulation becoming cluster members by $z=0$.}
\end{figure}

\begin{figure}
\includegraphics[width=1.\columnwidth, angle=0]{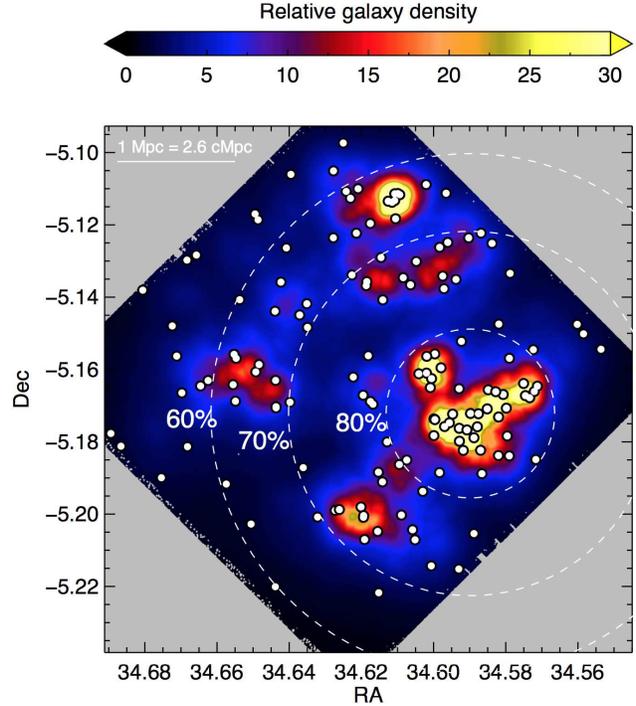}
\caption{\label{map_interloper} A map of \clustername\ with dashed circles marking where 80, 70 and 60\% of the galaxies are likely to become cluster members by $z=0$. White circles mark galaxies, and background colour scale indicates the relative galaxy density determined as the cumulative distance to the 5th nearest neighbour. }
\end{figure}

A major issue in photometric redshift surveys of galaxy clusters or protoclusters is the contamination level by line-of-sight interlopers. By sampling the Balmer and 4000\AA\ breaks of the protocluster galaxies, we were able to obtain precise photometric redshifts, and the contamination level due to photometric redshift uncertainty is only $12$\%. However some galaxies within the volume of the protocluster do not fall into the cluster by $z=0$ \citep{Muldrew2015,Contini2016}. Therefore not all galaxies that have similar redshifts as the main halo will fall into the cluster, and some of the \lq protocluster galaxies\rq\ selected in Section \ref{sec:selectingPCGal} will not be true cluster progenitors. 

We use the Planck-scaled Millennium Simulation with the semi-analytic model of \citet{Henriques2015} to determine which of \clustername 's galaxies will become cluster members by $z=0$. Using the 40 \clustername-like protoclusters at $z=1.61$ in the Millennium Simulation we select all galaxies within boxes of $10.2\times10.2\times34$\,cMpc volume around the main haloes, and then follow their evolution to determine if they become cluster members. 

In Fig.\,\ref{interloper_prob} we show the probability a galaxy will become a cluster member if it lies at a certain projected radius from the main halo. Although there is a large dispersion between radius and projected radius, there is still a strong correlation between projected radius and the probability a galaxy will fall into the cluster. Galaxies within 2\,cMpc of the main halo have more than 80\% chance of becoming cluster galaxies, whereas those that lie more than 6\,cMpc away have only a 60\% chance of making it into the cluster. \footnote{If the control field sample is 22\% more incomplete than the protocluster sample (see Section \ref{sec:selectingPCGal}) then the observations probe a protocluster depth of 41cMpc rather than 34cMpc. This results in the probabilities of Figs.\,\ref{interloper_prob} and \ref{map_interloper} decreasing by $\sim5$\%.}

The chance of becoming a cluster member also depends on whether the galaxy lies within a group or between the groups. In Fig.\,\ref{interloper_prob} we divide the galaxies into those which lie in groups more massive than the smallest group of \clustername, and those which lie in smaller groups or between the groups (labelled as intergroup galaxies). At all radii group galaxies are approximately 10\% more likely to  become cluster galaxies than intergroup galaxies, but for both subsets the probability of becoming a cluster member diminishes with increasing projected distance.

In Fig.\,\ref{map_interloper} we mark the probability that the  \clustername\ galaxies will become cluster members by $z=0$. Galaxies in group 1 are already cluster members. Galaxies in groups 3, 4 and 5 have more than $85$\% chance of becoming cluster members, whilst members of the massive group 2 have an 80\% probability, and even members of group 6, which lies 5.7\,cMpc from the main halo have a high ($>70$\%) chance of falling into the cluster.  

A large fraction of the observed intergroup galaxies are also likely to fall into the cluster. Although most potential cluster members of \clustername\ are intergroup galaxies at $z\sim1.6$, Fig.\,\ref{map_interloper} shows that the intergroup sample contains a higher level of contamination than the groups. The cleanest sample of true protocluster galaxies consists of those that reside in the groups surrounding a protocluster, but this sample is highly incomplete, and possibly biased due to environmental galaxy evolution processes occurring in the dense groups.

\section{Conclusions}
\label{conclusions}

We use $\sim1$\% precision photometric redshifts of \clustername\ to select a sample of protocluster galaxies that is both clean and complete enough to trace the wide-field structure of the protocluster. We obtain these high precision redshifts by observing the protocluster with narrow bands that tightly bracket the Balmer and 4000\AA\ breaks of the protocluster galaxies. 

We find that two structural features signify the presence of the protocluster: a large number of massive galaxy groups, and a high density of galaxies that lie between the groups. The groups are prominent features in maps of stellar mass density. We conclude that protoclusters may be more reliably identified as stellar mass overdensities rather than galaxy overdensities, which are more prone to line-of-sight contamination.

We show that future studies of protoclusters should avoid examining the protocluster using only one type of galaxy as this can severely bias our view of the protocluster. We have shown that some galaxy groups are entirely missed when studying the protocluster through only red or star forming galaxies, and the red galaxies do not locate the majority of the intergroup galaxies. Protoclusters are cradles of forming galaxies, but the formation of these galaxies is unevenly distributed. To view the whole structure and see the full complexity of the protocluster we must identify all the different types of protocluster galaxies. 

By observing a large fraction of the \clustername\ protocluster we have determined how much material is available to grow the cluster. Using cosmological simulations to identify protoclusters with the same structure as \clustername\ we estimate that it will grow into a $2.7^{+3.9}_{-1.7}\times 10^{14}$\Msun\ cluster by the present day. We mapped the evolutionary growth of \clustername\ and found that while its evolution is not atypical, the mass of the main halo of \clustername\ at $z=1.62$ is larger than 77\% of galaxy groups that end up with the same final mass. In comparison to other clusters with the same final mass, environmental quenching started earlier for a larger fraction of \clustername\ members.

We further use the simulations to assign a probability to each galaxy in the protocluster map of becoming a cluster member by $z=0$. The probability of becoming a cluster member rapidly diminishes with increasing projected distance. At the same radii, group galaxies are more likely to become cluster galaxies than the intergroup galaxies, and there is a very high probability that all 6 galaxy groups in \clustername\ will coalesce to form a cluster. The cleanest sample of cluster galaxy progenitors consists of those that reside in the groups within a couple of Mpc of the largest group, but this sample will be highly incomplete, and possibly biased due to environmental galaxy evolution processes occurring in the dense groups. 

We have demonstrated that the future evolutionary growth of a protocluster can be estimated from its structure.  By mapping the architectures of a large sample of protoclusters and clusters across $5>z>0$ we can place them in evolutionary sequences that describe how clusters form. Such sequences are powerful tools for studying how galaxies form and evolve in a dynamic environment. 

\section{Acknowledgments}
We sincerely thank the referee for providing useful and constructive comments which improved this paper. NAH acknowledges support from STFC through an Ernest Rutherford Fellowship. EAC acknowledges support from STFC. SIM acknowledges the support of the SFTC consolidated grant ST/K001000/1 to the astrophysics group at the University of Leicester. This work is based on observations made with ESO Telescopes at the La Silla Paranal Observatory under programme ID 089.A-0126.
\bibliographystyle{mn2e}\bibliography{References,mn-jour}
\label{lastpage}
\clearpage
\end{document}